\documentclass[fleqn,usenatbib]{mnras}

\usepackage{newtxtext,newtxmath}

\usepackage[T1]{fontenc}

\DeclareRobustCommand{\VAN}[3]{#2}
\let\VANthebibliography\thebibliography
\def\thebibliography{\DeclareRobustCommand{\VAN}[3]{##3}\VANthebibliography}

\newcommand\gaia{\textit{Gaia }}

\newcommand\Nclean{$91\,740~$}

\usepackage{graphicx}	
\usepackage{amsmath}	

\title[Validating Gaia SB1 Orbits]
{Gaia Spectroscopic Orbits Validated with LAMOST and GALAH Radial Velocities}

\author[D. Bashi et al.]{
D. Bashi,$^{1}$\thanks{E-mail: dolevbashi@gmail.com}
S. Shahaf,$^{2}$
T. Mazeh,$^{1}$
S. Faigler,$^{1}$
S. Dong,$^{3}$
K. El-Badry,$^{4,5,6}$
H.~W. Rix,$^{6}$
and A. Jorissen$^{7}$ 
\\
$^{1}$School of Physics and Astronomy, Tel Aviv University, Tel Aviv, 6997801, Israel\\
$^{2}$Department of Particle Physics and Astrophysics, Weizmann Institute of Science, Rehovot 7610001, Israel
\\
$^{3}$Kavli Institute for Astronomy and Astrophysics, Peking University, Yi He Yuan Road 5, Hai Dian District, Beijing 100871, People's Republic of China\\
$^{4}$Center for Astrophysics | Harvard \& Smithsonian, 60 Garden Street, Cambridge, MA 02138, USA\\
$^{5}$Harvard Society of Fellows, 78 Mount Auburn Street, Cambridge, MA 02138\\
$^{6}$Max-Planck Institute for Astronomy, Königstuhl 17, D-69117 Heidelberg, Germany\\
$^{7}$Institut d’Astronomie et d’Astrophysique, Université Libre de Bruxelles, ULB, CP 226, Boulevard du Triomphe, 1050 Brussels, Belgium}
\date{Accepted XXX. Received YYY; in original form ZZZ}

\pubyear{2022}

\begin{document}
\label{firstpage}
\pagerange{\pageref{firstpage}--\pageref{lastpage}}
\maketitle


\begin{abstract}

The recently published \gaia DR3 catalog of $181\,327$ spectroscopic binaries (SB) includes the Keplerian elements of each orbit but not the measured radial velocities (RVs) and their epochs. 
Instead, the catalog lists a few parameters that characterize the robustness of each solution.
In this work, we use two external sources to validate the orbits --- $17\,563$ LAMOST DR6 and $6\,018$ GALAH DR3 stars with measured RVs that have
\textit{Gaia}-SB orbits. 
We compare the expected RVs, based on the \gaia orbits, with the LAMOST and GALAH measurements. Finding some orbits that are inconsistent with these measurements, we constructed a function that estimates the probability of each of the \gaia orbits to be correct, using the published robust parameters. 
We devise a clean but still very large Gaia SB1 sample of \Nclean orbits.  The sample differs from the parent sample by the absence of --- physically unlikely and hence presumably spurious --- short-period binaries with high eccentricity. 
%
%
The clean SB1 sample 
offers the prospect of thorough statistical studies of the binary population after carefully modeling of the remaining selection effects. 

At a first look, two possible features emerge from the clean sample  --- 
a paucity of short-period binaries with low-mass primaries, which might be a result of some observational bias, and
a sub-sample of main-sequence binaries on circular orbits, probable evidence for circularization processes.

\end{abstract}




\section{Introduction}

The \gaia latest release of Non-Single Star catalogs 
\citep[][hereafter {\it NSS}]{NSS}
includes the orbits of $181\,327$ single-lined spectroscopic binaries (SB1), 
based on the radial velocities (RVs) obtained by the space-mission RVS spectrograph 
\citep[see also][]{RVS_I_22,RVS_II_22,DR3_Katz}. The sample (hereafter {\it NSS-SB1}) is much larger than any previously-known catalog of SB1s. For example, the SB9 catalog \citep[e.g.,][]{SB9} lists $\sim 4\,000$ orbits, while in a recent work based on the APOGEE project, \cite{Price-Whelan20} have identified $\sim 1\,000$ orbits.

By increasing the known SB1s by two orders of magnitudes, the new catalog is a gold-mine candidate for learning about the statistical features of short-period binaries, like the eccentricity-period relation \citep[e.g.,][]{Mazeh2008,jorissen09}, the frequency of binaries as a function of the  primary mass \citep[][]{raghavan10,troup16, moe17} and mass-ratio distributions \citep[e.g.,][]{mg92,boffin12, boffin15,shahaf17}. These features have profound implication on our understanding of binary formation and evolution \citep[e.g.,][]{verbunt95,bate97,bate02,harada21}, and therefore were intensively discussed in the past \citep[e.g.,][]{duchene13,shahaf19}, based on the relatively small samples available then. The field is now open for new detailed studies based on the much-larger \gaia-SB1 catalog. 

The \gaia catalog includes the Keplerian elements of the orbits, but not the \gaia RVs and their epochs themselves. 
Naturally, some erroneous orbital solutions are probably hidden in the catalog. 
For example, one can expect that a time series consisting of a small number of RV measurements with random observational noise could be fitted by a spurious short-period solution.

To help identify these erroneous cases, the catalog includes a few parameters that characterize the robustness of each orbit, like the \texttt{significance}, defined as the ratio between the primary semi amplitude $K$ and its uncertainty, and the \texttt{rv\_expected\_sig\_to\_noise}, defined as the signal-to-noise-ratio (SNR) of the \gaia RVS-spectra used.
As detailed by {\it NSS}, these parameters can be used to identify the more robust orbits and construct a sub-sample with a lower degree of false-orbit contamination. 

Such an exercise was performed by the authors of {\it NSS} themselves when they considered the eccentricity-period diagram. Indeed, the full sample includes very short-period binaries with high eccentricity (VSPHE), expected to be rare
\citep[e.g.,][]{meibom05,Mazeh2008,winn16,price-whelan18,terquem21,barker22}. 
When {\it NSS} considered only SB1s with 
\texttt{significance}$>40$, these VSPHE systems disappeared. 
Such a rejection of suspected binaries always comes with a price tag. For example,  the \texttt{significance}$>40$ threshold left a restricted sub-sample of only $28\,930$, smaller by a factor of $\sim6$ than the original one.

Quality assessment and validation using other data sets were also performed by {\it NSS} 
\citep[see \gaia DR3 documentation][]{rimoldini22}.\footnote{https://gea.esac.esa.int/archive/documentation/GDR3/pdf}
They presented a comparison between the \gaia DR3 orbital parameters and those from external catalogues --- SB9 \citep{SB9}; 
APOGEE \citep{Price-Whelan20} and WISE \citep{Petrosky21},   
with success rates of $\sim 80\%$.
As expected, when considered 
systems with orbital periods shorter than the RVS-observations time span, the recovery rate exceeds $90$\%.

In this work, we 
use two external sources of information --- the LAMOST DR6 
RV\footnote{https://dr6.lamost.org/} \citep{cui12} 
and the Galactic Archaeology
with HERMES (GALAH) DR3\footnote{https://www.galah-survey.org/dr3/the\_catalogues/} 
\citep{Buder21} surveys.
Instead of comparing orbital elements, we compare single RVs of $17\,563$ and $6\,018$ LAMOST and GALAH stars with the velocities expected by the \gaia orbits. 
\cite{DR3_Katz} used a similar approach to compare the whole \gaia single RVS catalog with a few RV external catalogs.

We could validate the \gaia orbit for most cases, but encountered about $10$--$20$\% inconsistent RVs. Using these data, we constructed a function that estimates the probability of each of the {\it NSS-SB1} orbit to be a false solution, based on the \gaia period and a few of the published robustness parameters. Choosing a working point that allows for $\sim10$\% contamination, we were able to select a 'clean' catalog, consisting of \Nclean orbits. 

Section~\ref{sec:comparison} compares the \gaia orbits with LAMOST and GALAH RVs. Section~\ref{sec:validation} suggests a clean SB1 sub-sample and presents some of its statistical aspects, and Section~\ref{sec:statistics} discusses two possible features of the sub-sample. We then briefly summarize and discuss our results in Section~\ref{sec:summary}.

\section{Comparing GAIA SB1 orbits
with LAMOST and GALAH RVs}
\label{sec:comparison}

\subsection{\gaia SB1s Cross match with LAMOST and GALAH}

We began by cross-matching the LAMOST DR6 sample of $5\,776\,260$ RVs with the \texttt{gaia\_DR3\_source} catalog, using a sky match of < $1$ arcsec. Next, we removed sources with missing estimates of the RV uncertainty and low signal-to-noise ratio (SNR) in LAMOST g 
(\texttt{snrg < $10$}) 
and i filters (\texttt{snri < $10$}), leaving us with $5\,146\,355$ LAMOST RVs associated with \gaia identified stars.   

To avoid cases where the LAMOST pipeline has mistakenly grouped different observed sources as the same source, we performed a few quality cuts for sources having more than one observation ($2\,091\,842$ RVs), using the LAMOST parameters $T_{\mathrm{eff}}$, $\log{g}$ and $\mathrm{[Fe/H]}$ which were derived for each exposure.  
We removed RVs that one or more of their LAMOST parameters deviated by more than $3\sigma$  from the distribution of this parameter, for the corresponding star.
This left us with  $5\,074\,002$ LAMOST RVs of $3\,850\,858$ different \gaia sources. 

\begin{figure}
\includegraphics[width=9cm]{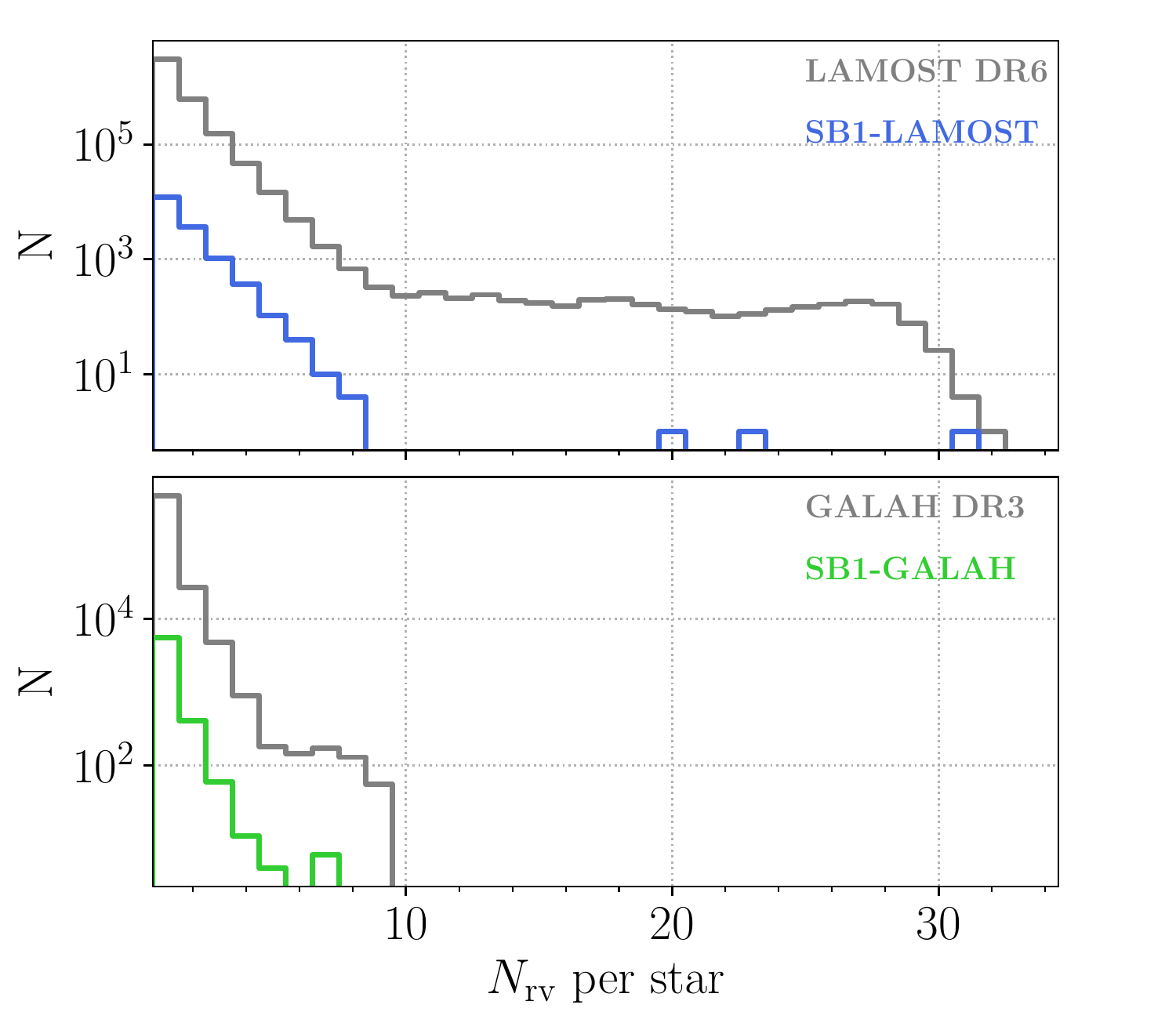}
    \caption{Histogram of RV measurements per star for LAMOST DR6 (upper panel) and GALAH DR3 (lower panel) sources, and for those that were found with \textit{Gaia} SB1 orbits (blue and green).}
    \label{fig:Nrvs LAMOST-SB1 Hist}
\end{figure}

We then cross matched this sample, based on \gaia source ID, with the \texttt{nss\_two\_body\_orbit} table of SB1's ($181\,327$ sources with \texttt{nss\_solution\_type = SB1}).
We found $17\,436$ systems in common, of which $12\,230$ objects had a single LAMOST RV measurement, as shown in Fig.~\ref{fig:Nrvs LAMOST-SB1 Hist}.
Three sources
had $20$, $23$ and $31$ observations, for which 
we compared the independently-derived LAMOST orbits with those of {\it NSS-SB1} (see subsection \ref{three-systems}). 

A similar cross-match was performed with the GALAH DR3 \citep{Buder21} RVs. 
We considered only measurements with available non gravitational-redshift-corrected RV (\texttt{rv\_nogr\_obst}),
resulting in a sample of $562\,887$ RV measurements of $519\,668$ distinct sources.
We found $6\,018$ stars in common with the \gaia orbits, 
$5\,552$ of which had a single GALAH-RV measurement,
as seen in Fig.~\ref{fig:Nrvs LAMOST-SB1 Hist}.

\begin{figure*}
	\includegraphics[width=14 cm]{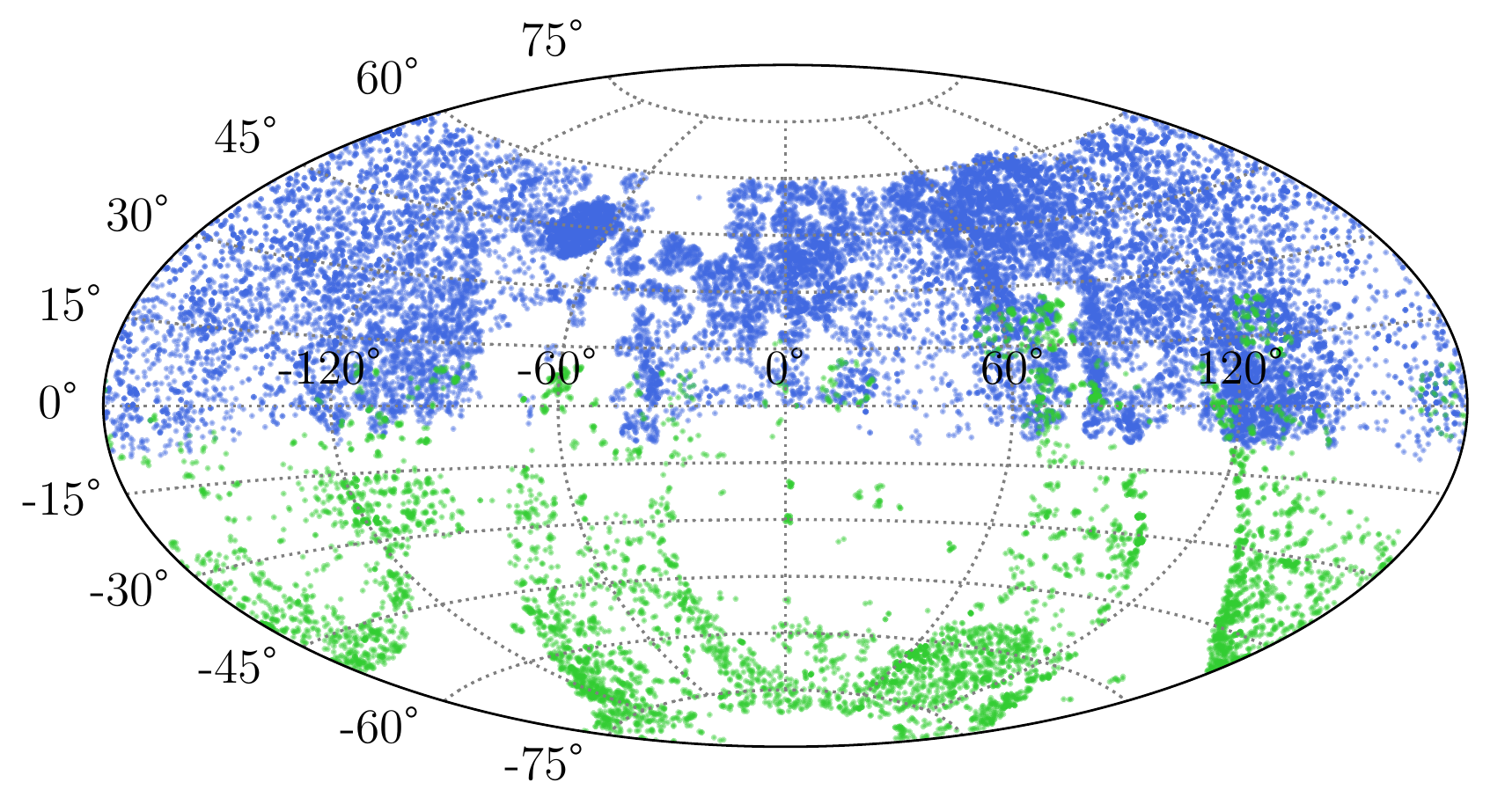}
\caption{Sky projection of $17\,436$
LAMOST (blue) and  $6\,018$ GALAH (green) observed stars with \gaia orbits.}
    \label{fig:SKY}
\end{figure*}

Density distributions of the LAMOST and GALAH systems that are in common with the \gaia SB1 binaries are plotted in equatorial coordinates in Fig.~\ref{fig:SKY}.
One can see that the LAMOST sources (marked in blue) are spread over the northern hemisphere while the GALAH sources (marked in green) are located in the south. $357$ sources are common to the two samples.

\subsection{Three LAMOST orbits}
\label{three-systems}

The best way of validating the \gaia orbits, as was done by {\it NSS} themselves, is to compare the \gaia Keplerian orbits with the other surveys' orbits. Unfortunately, this can not be done with the LAMOST or GALAH data sets as these surveys, by their nature, did not accumulate enough RVs for almost all of their observed stars.
Nevertheless, as can be seen in 
Fig.~\ref{fig:Nrvs LAMOST-SB1 Hist}, there are three LAMOST exceptions for which we could derive independent orbital solutions, as detailed in Table~\ref{tab:LASMOT+20sources} of Appendix~\ref{sec:three-solutions}. Furthermore, we plotted the LAMOST RVs and \gaia orbits, phase folded by the \gaia period, by calculating the  LAMOST phases (including uncertainties) at the \gaia orbits (Figs.~\ref{fig:GaiaModel_3425964192178375680}--\ref{fig:GaiaModel_3376949338201658112}) 

We could verify the \gaia orbit of only one of the three systems --- \textit{Gaia} DR3 $3425964192178375680$. The LAMOST RVs of another system, \textit{Gaia} DR3 $3376949338201658112$, (see Fig.~\ref{fig:GaiaModel_3376949338201658112}) did not show any variation, and the third system,
\textit{Gaia} DR3 $3425556586900263424$,
yielded a similar but different-period solution. 

Although a very small number, the three examples demonstrate two major points.
First, some of the \gaia orbits are probably false with no real variation. However, even for the case of \textit{Gaia} DR3 $3376949338201658112$, a few LAMOST RVs are consistent with the \gaia predicted velocities. Those are close to $\gamma$, the center-of-mass velocity of the orbit. Consequently, one can not verify the \gaia orbit with one LAMOST RV, unless that velocity is consistent with the orbit and significantly different from the \gaia $\gamma$ (see below). 
Second, the orbital period of \textit{Gaia} DR3 $3425556586900263424$ is wrong. Nevertheless, most, if not all, the LAMOST RVs are within $1$--$2\sigma$ away from the \gaia expected velocities. Again, one LAMOST RV consistent with the \gaia orbit is no proof that all elements of the orbit are valid.

\subsection{Deriving the difference between the \textit{Gaia}-expected RVs and the LAMOST and GALAH measurements}


To compare the LAMOST velocities with the \gaia orbits, we used the reported \textit{Gaia} Keplerian parameters:
        \begin{itemize}
        \item[--] $P$ (\texttt{period}): orbital period
        \item[--] $K$ (\texttt{semi\_amplitude\_primary}): semi-amplitude of the radial velocity curve related to the first component
        \item[--] $e$ (\texttt{eccentricity}): eccentricity of the orbit        
        \item[--] $t_p$ (\texttt{t\_periastron}): periastron epoch
        \item[--] $\omega$ (\texttt{arg\_periastron}): argument of periastron
        \item[--] $\gamma$ (\texttt{center\_of\_mass\_velocity}): velocity of the centre of mass.
        \end{itemize}  
We used these parameters to derive the expected RV --- $RV_{\rm SB1}$, at the LAMOST epoch, MJD $t_L$, and compared it 
to the LAMOST observed RV, $RV_{\rm L}$, with the difference
%
\begin{equation}
   \mathcal{D}_{\rm SB1,L} = RV_{\rm SB1} - RV_{\rm L} \ . 
\end{equation}

To derive the uncertainty of $RV_{\rm SB1}$, we used the reported \gaia-Keplerian-parameters correlation matrix, \texttt{corr\_vec}, to produce
$1\,000$ random draws of Keplerian parameters, excluding cases with negative or larger-than-one eccentricity value.  
The standard deviation (STD) of these $1\,000$ velocities was considered as the $RV_{\rm SB1}$ uncertainty --- $\sigma[RV_{\rm SB1}]$.  The uncertainty of the difference was then derived by quadrature
of the uncertainties of \gaia-expected and LAMOST velocities:

\begin{equation}
    \sigma[\mathcal{D}_{\rm SB1,L}] =\sqrt{\sigma [RV_{SB1}]^2 + \sigma[ RV_{L}]^2} \ .
\end{equation}
%
We denote the difference $\mathcal{D}_{\rm SB1,L}$ in units of its own uncertainty as
\begin{equation}
    \mathcal{D}_{\rm SB1,L}^{\sigma}=
    \frac{\mathcal{D}_{\rm SB1,L}}
         {\sigma[\mathcal{D}_{\rm SB1,L}]}\ . 
\end{equation}
A similar procedure was performed with the GALAH RVs, 
obtaining $\mathcal{D}_{\rm SB1,G}$, $\sigma[\mathcal{D}_{\rm SB1,G}]$ and 
$\mathcal{D}_{\rm SB1,G}^{\sigma}$.

\begin{figure*}
\includegraphics[width=18cm]{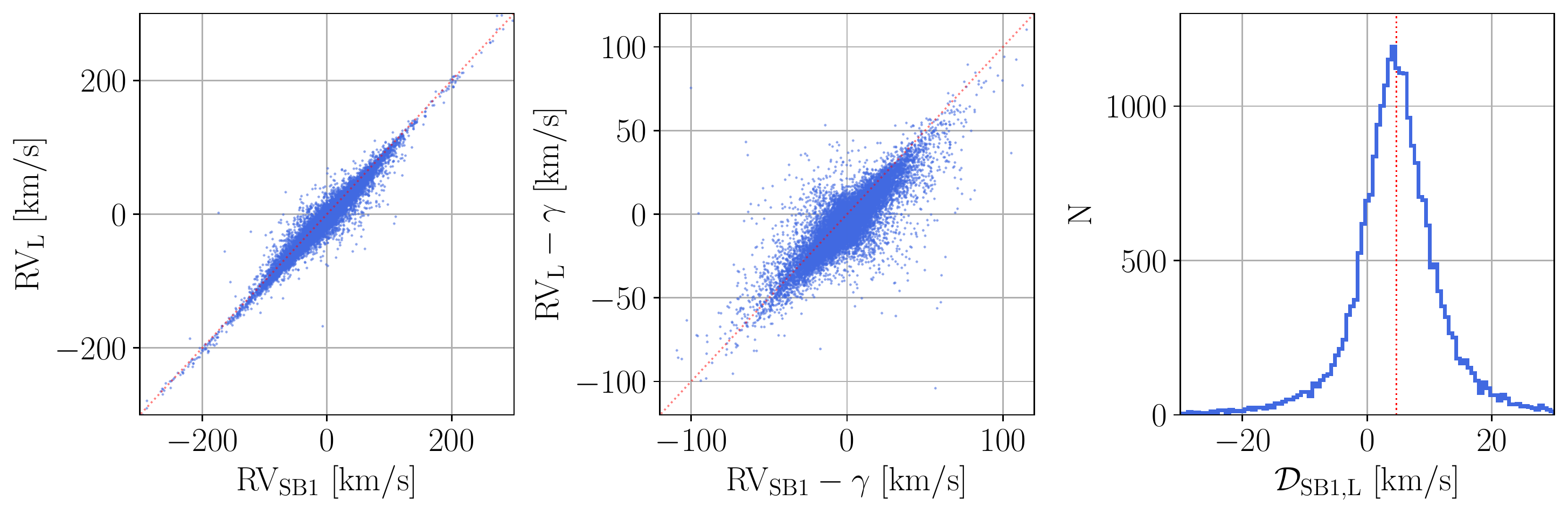}
\includegraphics[width=18cm]{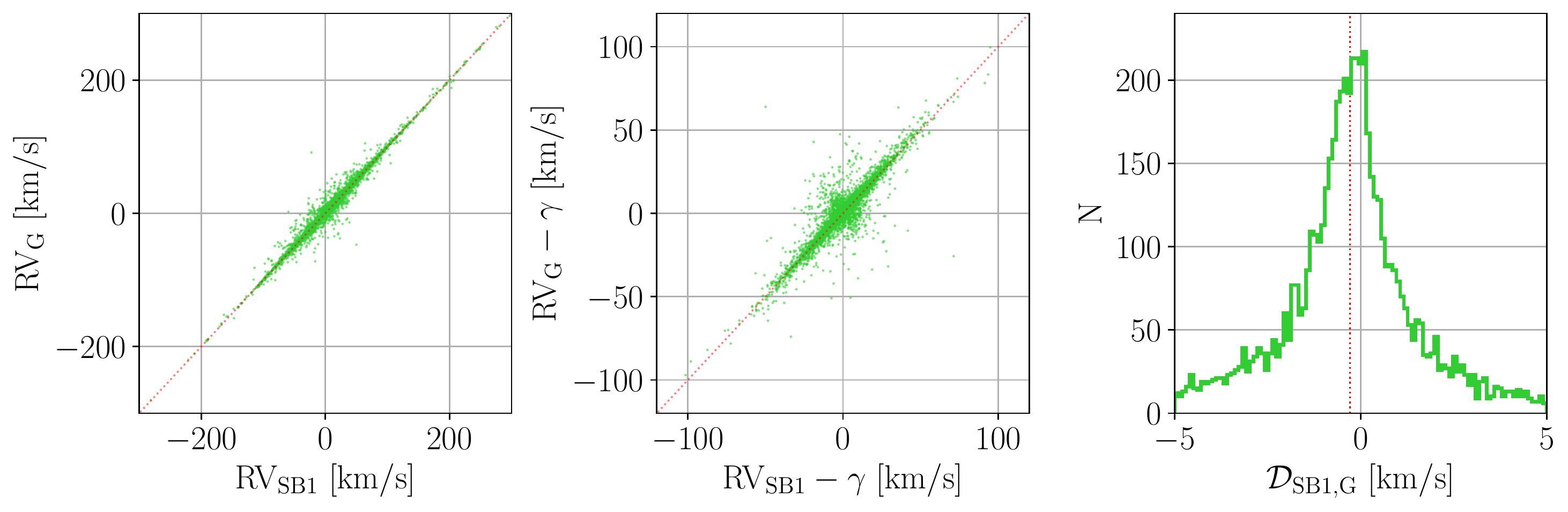}
    \caption{
    Differences between the LAMOST and GALAH RVs and the \gaia orbits. 
    The left two panels present the RVs reported by LAMOST (upper panel; blue) and GALAH (bottom panel; green) as a function of the \gaia expected RVs. 
    The middle panels show those velocities after subtracting \gaia $\gamma$, the estimated center-of-mass velocity of each binary.
    The right panels show histograms of the difference between \textit{Gaia} RV model and the LAMOST DR6 (upper) and GALAH DR3 (lower) measurements. Red dashed vertical lines mark the median values, suggesting offsets of $4.7$ km/s and $-0.28$ km/s of LAMOST and GALAH RVs, respectively, relative to the \gaia ones. }
    \label{fig:SB1_LAMOST_GALAH_rv_diff}
\end{figure*}

The differences between the LAMOST and GALAH RVs and the \gaia orbits are plotted in Fig.~\ref{fig:SB1_LAMOST_GALAH_rv_diff}. 
The left two panels present the RVs reported by LAMOST (upper panel; blue) and GALAH (bottom panel; green) as a function of the \gaia expected RVs. These panels show a general agreement between the orbits and the observed RVs. The middle panels show those velocities after subtracting the \gaia $\gamma$ velocity of each binary, displaying large differences for some of the systems. Histograms of the RV differences are plotted in the right two panels.
The figure does not include the largest differences, in order to present the behaviour around zero difference. The median of the LAMOST-\gaia differences, denoted by a red dashed line, is $4.74$ km/s, suggesting a LAMOST RV zero-point offset. Similar shifts were noticed by \citet[][based on LAMOST DR3]{Anguiano18} and by \citet[][based on LAMOST DR7]{DR3_Katz}. For GALAH, a $-0.28$ km/s shift is emerging.

\begin{figure*}
	\includegraphics[width=14cm]{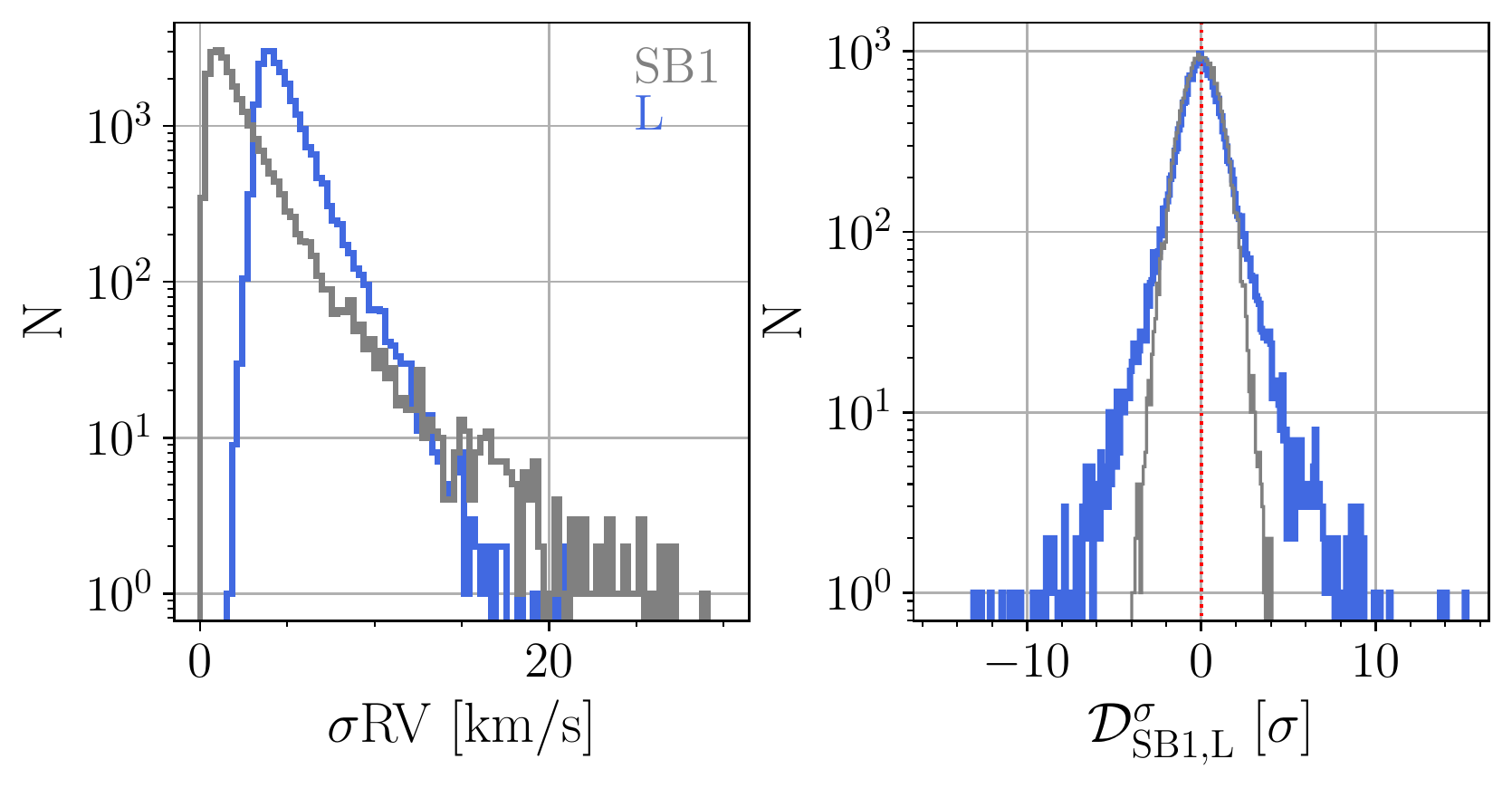}
		\includegraphics[width=14cm]{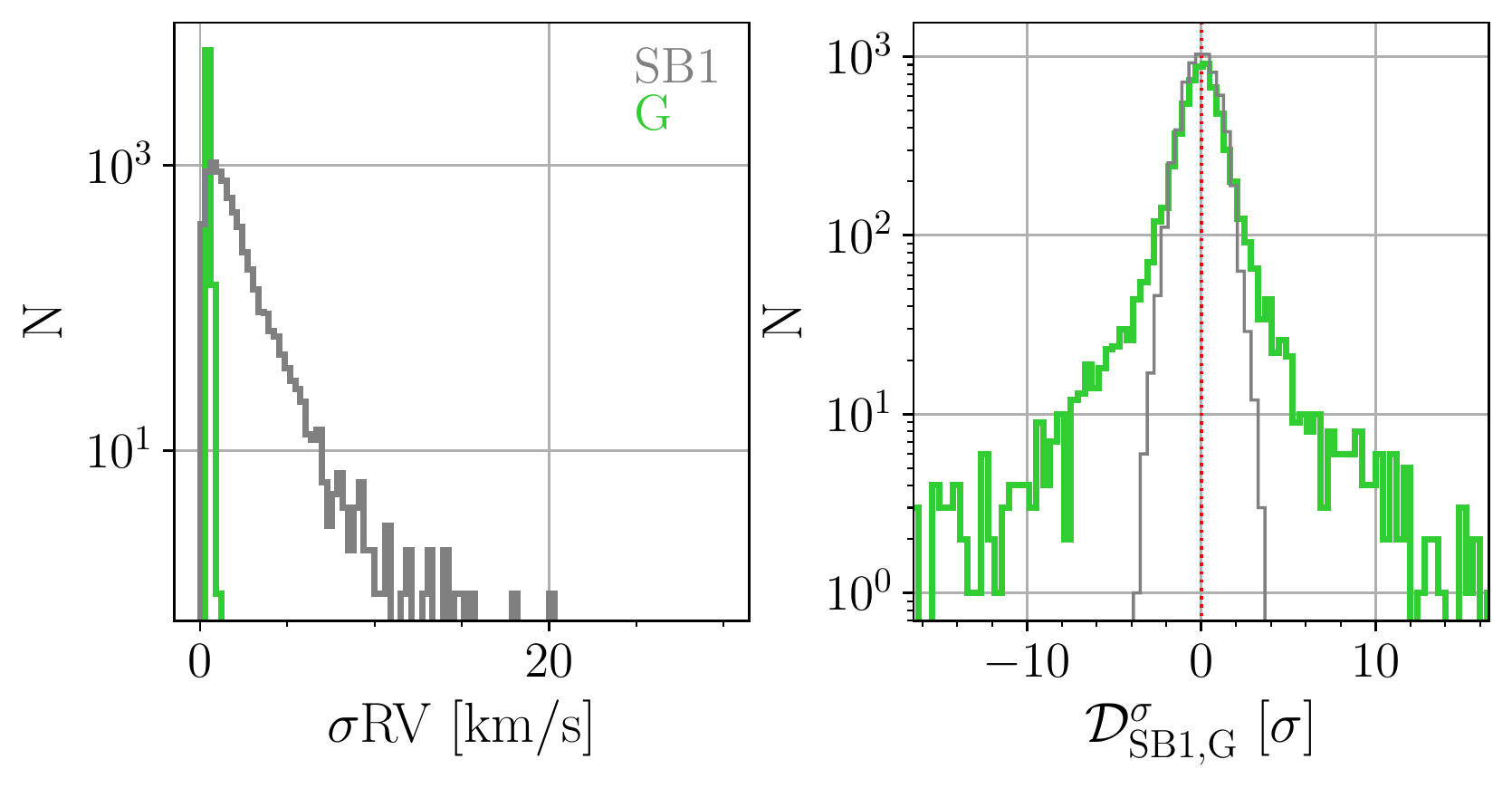}
    \caption{Left panels show histograms of RV uncertainties of the LAMOST (upper left, blue) and GALAH (bottom left, green) measurements. The displayed uncertainties are after a $0.5$ km/s addition in quadrature to the original GALAH uncertainties (see text). Also plotted histograms of the \gaia uncertainties of those orbits (grey). Right panels present histograms of the differences, in their uncertainty units, between \textit{Gaia} models and $24\,757$ LAMOST (upper right, blue) and $6\,628$ GALAH (lower right, green) measurements. 
    Right-panels grey curves show simulated normal distributions with a mean of $0$ and STD of $1$. The outliers-immune STD of the normalized differences, based on Median-Absolute-Deviation (MAD), is  $1.07$ for LAMOST and $1.20$ for GALAH.}
    \label{fig:SB1_LAMOST_rv_dif_fix}
\end{figure*}

Fig.~\ref{fig:SB1_LAMOST_rv_dif_fix} shows in its two left panels histograms of the adopted uncertainties of the LAMOST and GALAH RVs, together with the uncertainties of the corresponding \textit{Gaia}-expected velocities. 
Based on \citet{Buder21}, we opted to add in quadrature $0.5$ km/s to the reported uncertainty of the GALAH RVs. 
This relatively small uncertainty
is due to the high spectral resolution of the GALAH spectra,  $R \sim 28\,000$, compared with that of \gaia, $ \sim 11\,500$ and that of 
LAMOST LRS, which is only $ \sim 1\,800$. 

The right two panels of the figure show histograms of the RV differences (after correcting for the LAMOST and GALAH zero points), in units of their uncertainties, $\sigma[\mathcal{D}_{\rm SB1,L}]$ and  $\sigma[\mathcal{D}_{\rm SB1,G}]$. We also plot a standard normal distribution expected for the same number of measurements.

The histograms present large wings that indicate a substantial number of non-Gaussian differences.
The fact that the significant differences appear in both the LAMOST and GALAH histograms suggests that most of them stem from erroneous \gaia solutions, which is the main interest of this work, although we cannot rule out some cases for which the LAMOST or GALAH RVs are wrong. We show in Section \ref{sec:validation} that the significant RV differences occur in specific \gaia orbits, corroborating the notion that the differences are due to mistaken orbits.

\section{Validation the \gaia orbits}
\label{sec:validation}

\subsection{Identification of validated orbits}

To distinguish between true and false \gaia orbits,
we compare the LAMOST and GALAH RVs with two competing models:

(i) The  \gaia orbits, where we use  $\mathcal{D}_{\rm SB1,L}^{\sigma}$ and  $\mathcal{D}_{\rm SB1,G}^{\sigma}$ to quantify the agreement between the RVs and the expected \gaia RVs in LAMOST and GALAH epoch.

(ii) No RV modulation at all; the \gaia stellar velocity is simply the $\gamma$ velocity of the orbit. 
For this model, we construct the differences for LAMOST RVs as
%
\begin{equation}
   \mathcal{D}_{\rm \gamma,L} = \gamma - RV_{\rm L} \ , 
\end{equation}
%
\begin{equation}
    \sigma[\mathcal{D}_{\rm \gamma,L}] =
    \sqrt{\sigma[\gamma]^2 + \sigma[RV_{L}]^2} \ ,
\end{equation}
and
\begin{equation}
    \mathcal{D}_{\rm \gamma,L}^{\sigma}=
    \frac{\mathcal{D}_{\rm \gamma,L}}
    {\sigma[\mathcal{D}_{\rm \gamma,L}]}\ . 
\end{equation}
Similar differences were derived for the GALAH RVs:\\ $\mathcal{D}_{\rm \gamma,G}^{\sigma}=
\mathcal{D}_{\rm \gamma,G}/\sigma[\mathcal{D}_{\rm \gamma,G}]$.

\begin{figure*}
	\includegraphics[width=8.8cm]{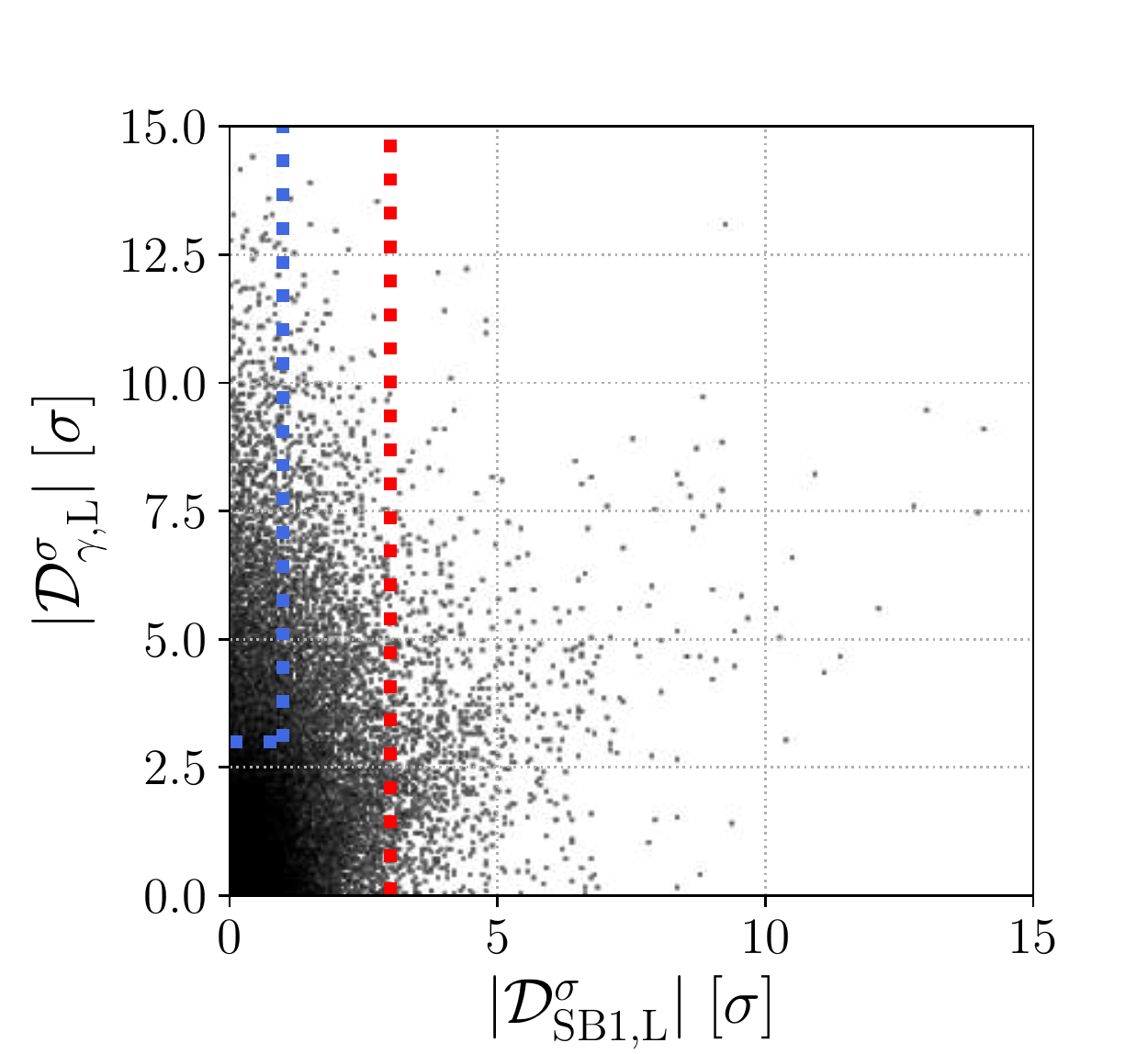}
		\includegraphics[width=8.8cm]{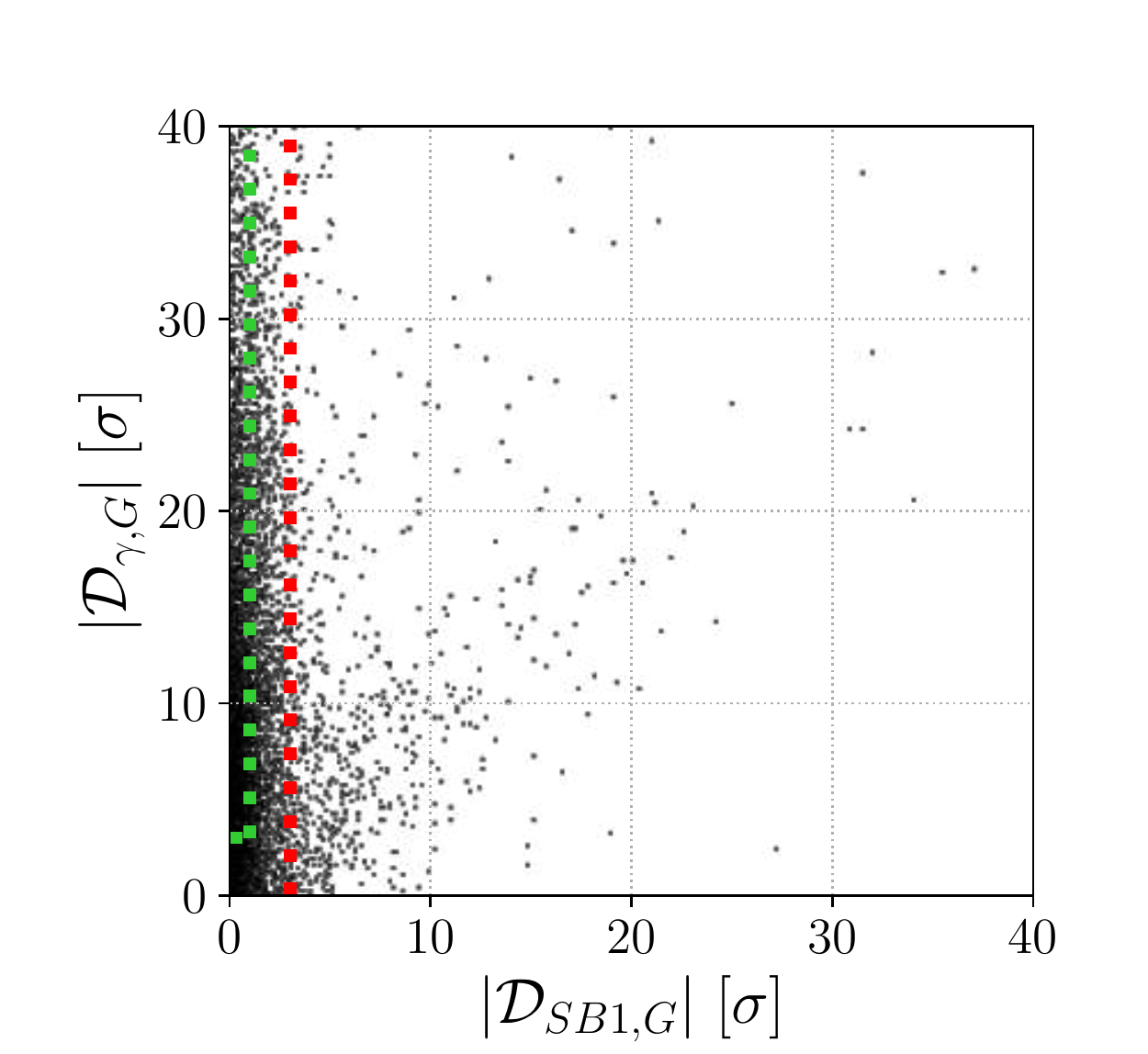}
\caption{Differences (in units of their uncertainties) between the LAMOST (left) or GALAH (right) RVs and two competing \gaia models --- constant RV and {\it NSS-SB1} model.  Points inside the blue or green rectangle ($2\,445$ or $2\,777$) are sources where the LAMOST or GALAH RVs agree, by up to $1 \sigma$, with the {\it NSS-SB1} model and disagree, by over $3 \sigma$, with the constant model. Points right to the red dashed lines ($975$ or $666$) are sources where the LAMOST or GALAH RVs are inconsistent, by over $3 \sigma$, with the {\it NSS-SB1} model. The rest of the black points in the background are inconclusive sources.}
    \label{fig:SB1_Sigmas}
\end{figure*}

Fig.~\ref{fig:SB1_Sigmas} displays the differences relative to the two models, in units of the corresponding uncertainties, for both the LAMOST and GALAH samples.

Accordingly, we divided our sample of \gaia SB1-LAMOST/GALAH sources into three groups:
\begin{itemize}
    \item[--] $2\,445$ and $2\,777$ systems for which the radial velocity predicted by the \gaia orbits fits well with the LAMOST and GALAH RVs, respectively, and are {\it not} consistent (by more than $1\sigma$) with the $\gamma$ velocity. 
    We consider these orbits as validated by LAMOST and GALAH RVs. 
    \item[--] $975$ and $666$ orbits for which at least one of the corresponding LAMOST and GALAH RVs are  more than $3\sigma$ away from the \gaia orbits,
    considered as refuted by LAMOST and GALAH. 
    \item[--] All the rest are systems for which the LAMOST and GALAH RVs are within $3\sigma$ from both models or between $1$--$3\sigma$ away from the \gaia orbits,
    and therefore their RVs are inconclusive.
   \end{itemize}

The separations between the different groups are marked in the figure.

\begin{figure*}
	\includegraphics[width=19cm]{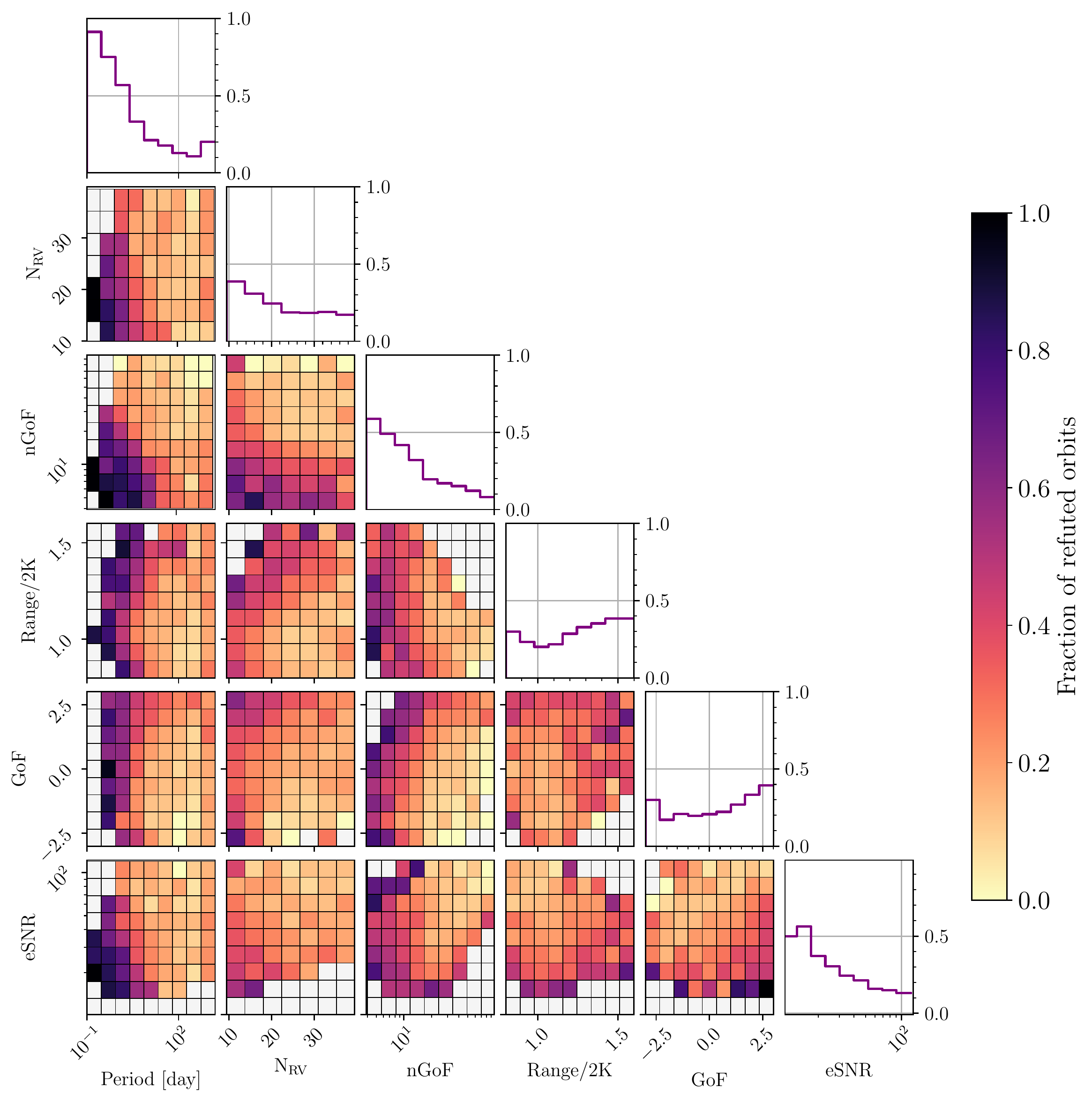}
    \caption{Corner plot of the relative number of refuted orbits as a function of six published  parameters: Period, N$_{\rm RV}$, nGoF, 
    Range/2K, GoF, and eSNR. Each 2D plot is divided into bins, with colours representing the fraction of refuted orbits in that bin. White bins have less than $5$ orbits, considered as not having enough information.}
    \label{fig:SB1_CornerRatios}
\end{figure*}

\subsection{The robust parameters and the validated orbits}

We now try identifying the \gaia robust parameters of the SB1 orbits that can identify low-quality solutions, 

such as short-period solutions with a small number of points, or low-SNR solutions. 

We consider a few of the parameters listed at the \texttt{nss\_two\_body\_orbit} and the \texttt{gaia\_source} tables:

(i) \texttt{Period} --- The SB1 period (Period).

(ii) \texttt{rv\_n\_good\_obs\_primary} --- Total number of epoch RVs actually used for the primary in solving the SB1 model (N$_{\rm RV}$).

(iii) \texttt{rv\_renormalised\_gof} --- An empirical value defined by the \gaia team, that compares the scatter of the RVs relative to the obtained solution with the typical RV uncertainty of stars with similar  \texttt{rv\_template\_teff} and \texttt{grvs\_mag}. As part of the compilation of the NSS table, only sources with \texttt{rv\_renormalised\_gof} > $4$ were considered (nGoF).

(iv) \texttt{rv\_amplitude\_robust}/$2\times$\texttt{semi\_amplitude\_primary} --- The total range of the observed RVs, given by the {\it NSS-SB1} catalog, divided by twice the semi amplitude of the SB1 model. This parameter estimates the coverage of observed RVs of the orbital RV range (Range/2K). 

(v) \texttt{goodness\_of\_fit} --- Goodness-of-fit statistic of the solution. This is the ‘Gaussianized chi-square’ \citep{rimoldini22} cube root transformation), which for good fits should approximately follow a normal distribution with zero mean value and unit standard deviation (GoF).

(vi) \texttt{rv\_expected\_sig\_to\_noise} --- Expected signal to noise ratio \citep{rimoldini22} in the spectra used to obtain the radial velocity (eSNR).

Other parameters such as the \texttt{conf\_spectro\_period} (i.e. the probability of the period not being due to Gaussian white noise) 

did not yield any clear trend. 
Attempting to use a smaller set of parameters resulted in lower performance of our classifier (Section \ref{subsec:classifier}). 
For example, a simpler cut with just a combination of Period and N$_{\rm RV}$ was inferior by $10-15\%$ to our adopted classifier. 

For each pair of these parameters, we derived a 2D binned map of the ratio of false orbits to the sum of the false and true orbits, presented in Fig.~\ref{fig:SB1_CornerRatios} in a corner-plot form. Obviously, the figure that presents the fraction of the refuted orbits depends on the separation between the false and valid orbits we adopted for Fig.~\ref{fig:SB1_Sigmas}.

Fig.~\ref{fig:SB1_CornerRatios} suggests that all six robust parameters have a significant predictive power.
In particular, the relative number of refuted orbits is high for short-period binaries, either with small nGoF, eSNR, or N$_{\rm RV}$ parameters. 

We note that the tendency for false orbits revealed by Fig.~\ref{fig:SB1_CornerRatios} is common to the LAMOST and GALAH RVs in all regions of the parameter space. One apparent exception is the long-period  orbits, for which the LAMOST fraction of RVs inconsistent with the \gaia orbits is larger than the corresponding GALAH RVs. The cause for this small difference is not clear and might be due to the larger fraction of giants in the long-period orbits.


\begin{table}
\caption{
Best-fitted coefficients of the 
Logistic-regression classification.
\label{tab:LR_coeff}}
\begin{center}
\begin{tabular}{l l r }
\hline
Coefficient & Parameter name & Value \\
\hline
$b_0$ &  & $3.8791$ \\
$b_1$ & $\mathrm{log}_{10}\mathrm{(Period)}$ & $-0.6062$  \\
$b_2$ & N$_{\rm RV}$& $-0.041$\\
$b_3$ & $\mathrm{log}_{10}\mathrm{(nGoF)}$ & $-2.9939$  \\
$b_4$ & Range/2K& $0.0153$  \\
$b_5$ & GoF & $ 0.4038$ \\
$b_6$ & $\mathrm{log}_{10}\mathrm{(eSNR)}$ & $0.9734$\\

\hline
\end{tabular}
\end{center}
\end{table}

\begin{figure}
	\includegraphics[width=9cm]{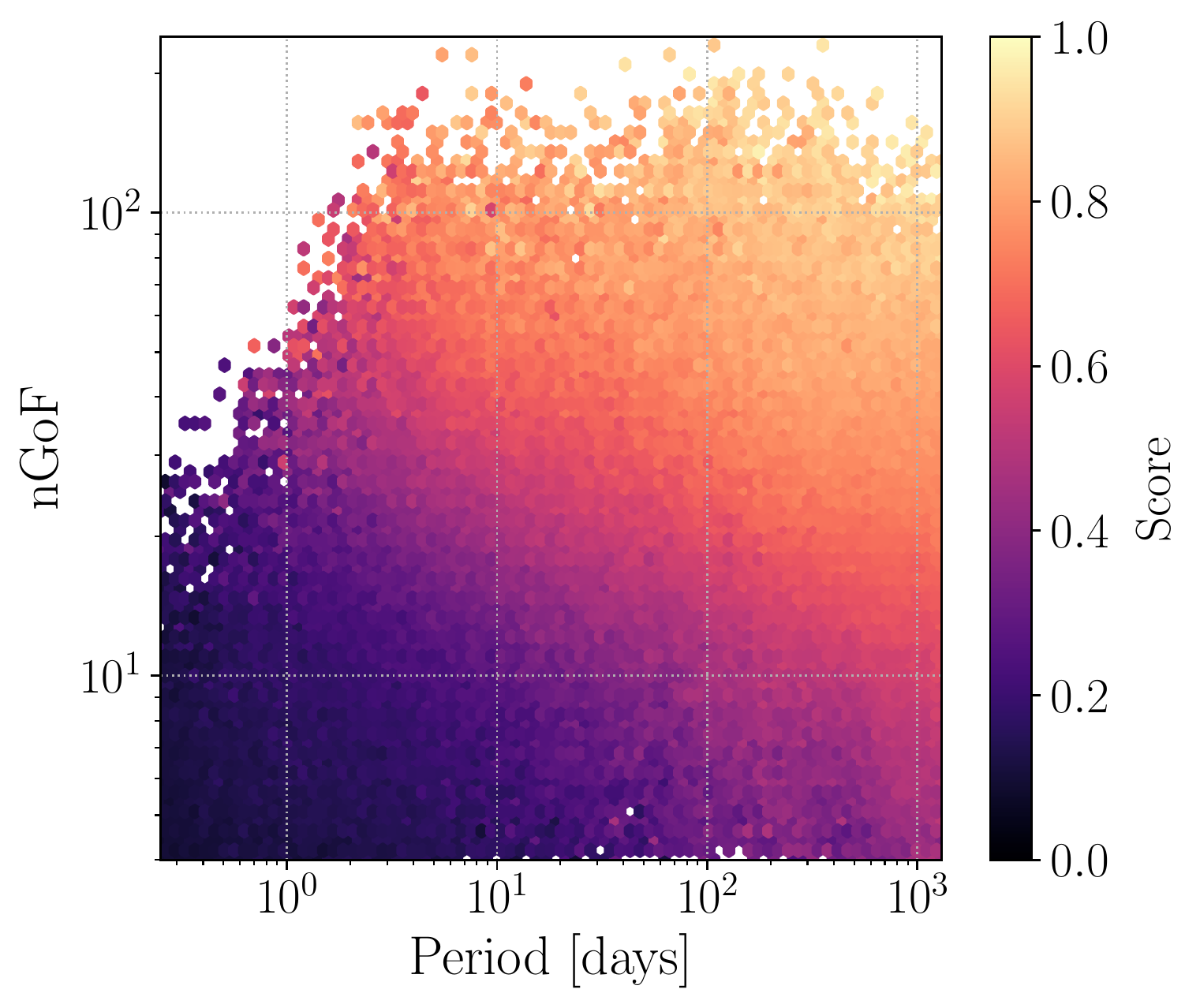}
    \caption{Mean score as a function of the period and nGoF, presented as 2D binned diagram, colour-coded by the mean logistic-function score at each bin.}
    \label{fig:per_nGOF_allscores}
\end{figure}

\begin{figure}
	\includegraphics[width=9cm]{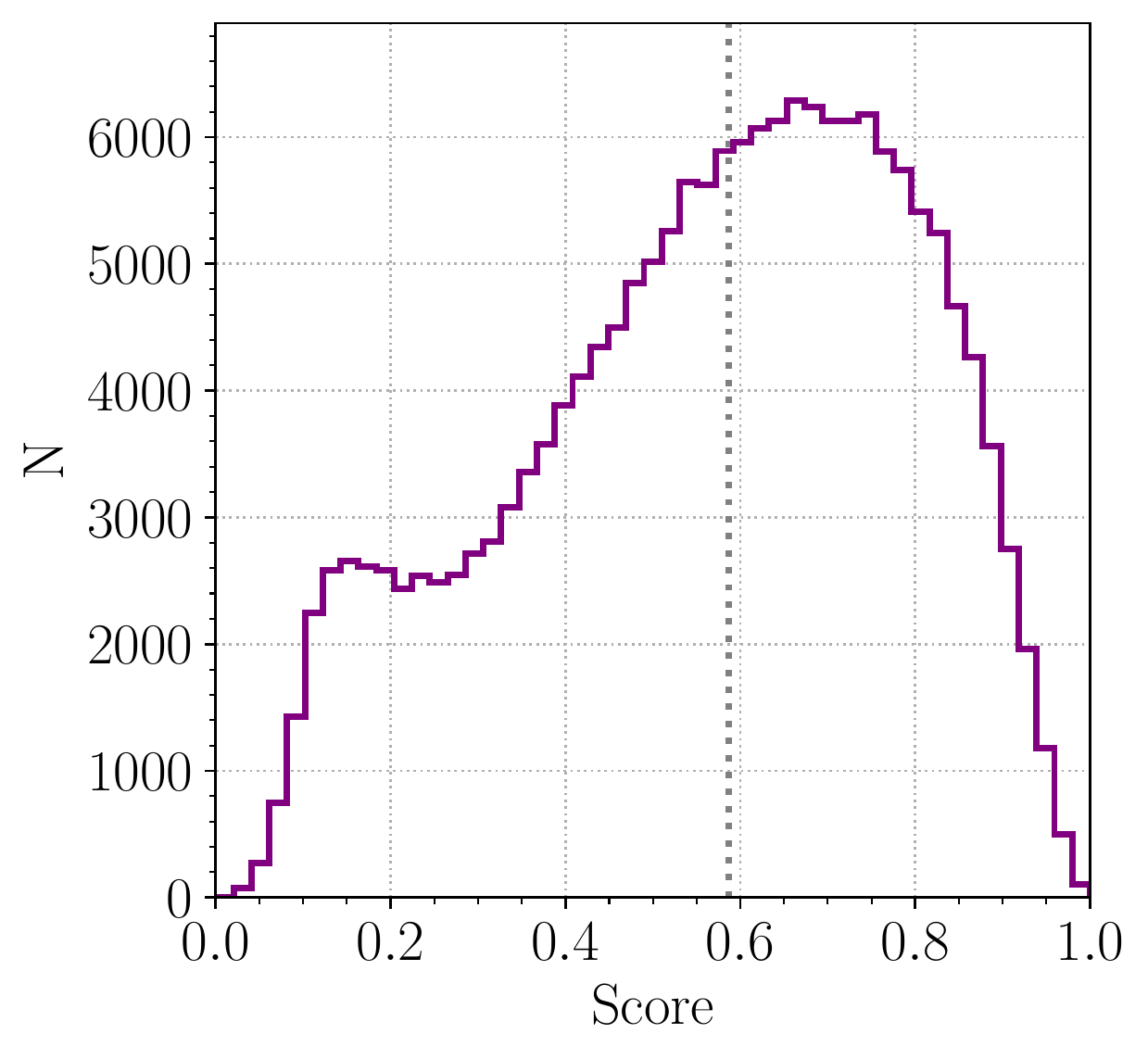}
    \caption{Histogram of logistic-regression scores of the full {\it NSS-SB1} sample. Low-score systems are probably false. Gray vertical line marks our selected score limit of $0.587$.}
    \label{fig:scores_hist}
\end{figure}

\subsection{Probability Function to identify the real \gaia orbits}
\label{subsec:classifier}

Following the qualitative trends presented in 

Fig.~\ref{fig:SB1_CornerRatios}, we built a statistical model using logistic regression
\citep{Bishop2006}\footnote{sklearn.linear\_model.LogisticRegression module in python} to estimate the probability that a \gaia SB1 solution is real, using the two groups of 
Fig.~\ref{fig:SB1_Sigmas}.
For orbits common to both LAMOST and GALAH, we chose to list them as refuted orbits when one or more of the RVs met our $3\sigma$ away from the \gaia orbit model criteria. We then built a training set sample, composed of $70\%$ of the sample, and a testing set (the other $30\%$ of the sample), and used a \texttt{sigmoid} function, commonly used in the logistic-regression framework, to find the best $N$-dimensional plane that separates the two groups. 
Specifically, the function derived is:
\begin{equation}
S(z) = \frac{1}{1+e^{\textbf{z}}} ,
\label{eq:logit}
\end{equation}
where $\textbf{z} = b_0 + b_1z_1 + b_2z_2 + b_3z_3 + b_4z_4 + b_5z_4 + b_6z_6$ are the classification best fitted coefficients, listed in Table~\ref{tab:LR_coeff}.

The derived logistic function assigns a score value between $0$ and $1$ that estimates the validity for all orbits of {\it NSS-SB1}. These values are released in the supplement of this paper, for the community to use.
Again, the derived function depends on the separation between the false and valid orbits we adopted. 

Fig.~\ref{fig:per_nGOF_allscores} shows the distribution of the score in the period-nGoF plane, with color code based on mean scores in each bin, while 
Fig.~\ref{fig:scores_hist} presents the distribution of the derived scores. A clear trend is evident in 
Fig.~\ref{fig:per_nGOF_allscores} that resembles the shape seen in the associate diagram of Fig.~\ref{fig:SB1_CornerRatios}.

\newpage
\subsection{The Clean Sample}

\begin{figure}
	\includegraphics[width=8cm]{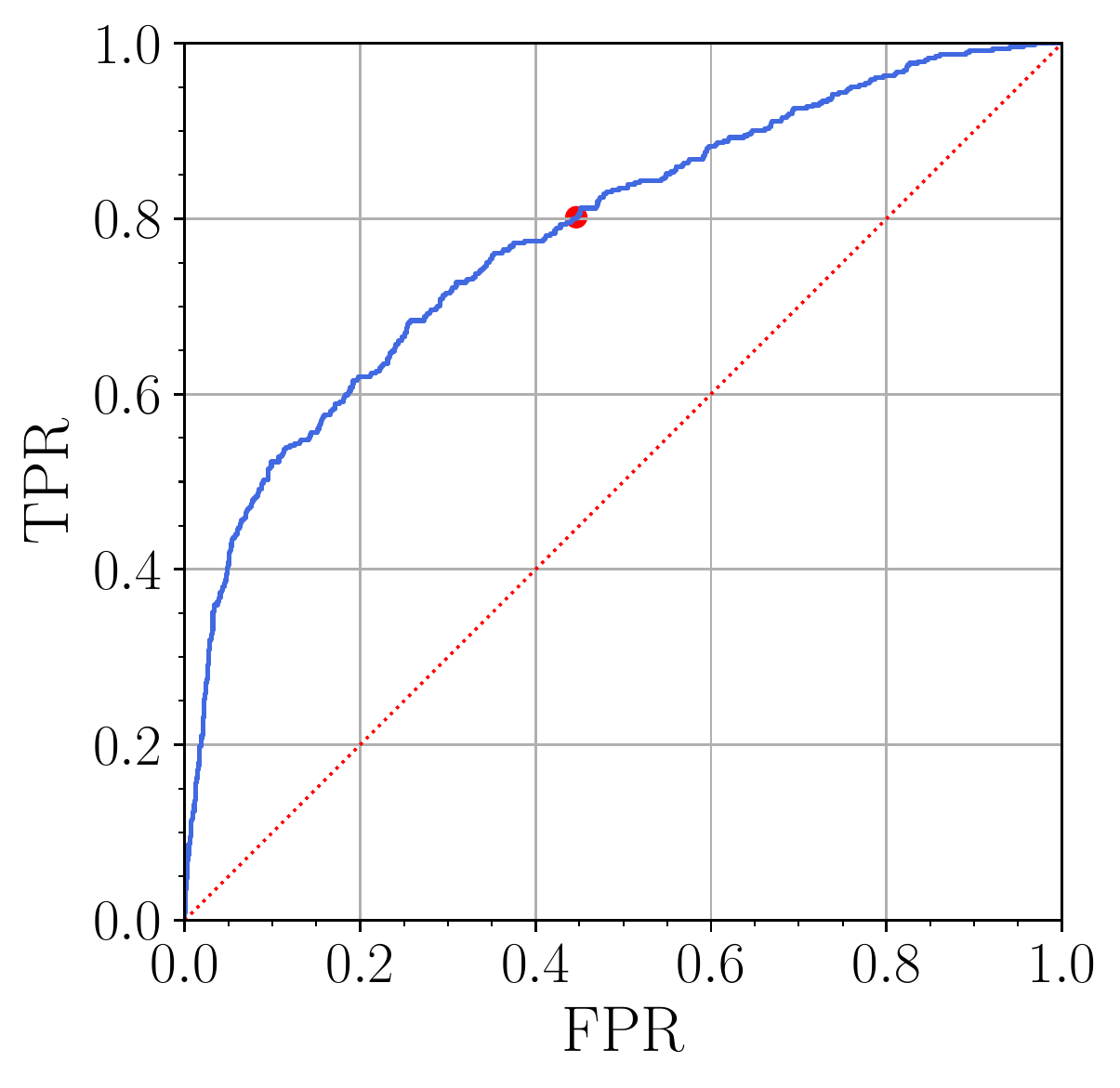}
    \caption{ROC curve and our selected score limit of $0.587$ (red point), aimed to set a True Positive Rate (TPR) of $0.80$. The Area Under the Curve (AUC) is $0.78$, clearly above the $0.5$ value expected for a random-model classifier, marked by the red dashed line.}
    \label{fig:ROC}
\end{figure}

To characterize the diagnostic ability of our binary 
classification we used a ROC (Receiver Operating Characteristic) curve \citep{fawcett06} presented in Fig.~\ref{fig:ROC}. A ROC graph is a technique for visualizing, organizing, and selecting classifiers based on their performance.  ROC curve is created by plotting the estimated True Positive Rate (TPR) against the estimated False Positive Rate (FPR) at various threshold settings.  The TPR (sensitivity) is a measure of the fraction of false orbits that are identified as such.

Adopting a score limit of $0.587$, with a sensitivity of $80\%$, implies that our 'clean' sample, with \Nclean orbits, might include $20\%$ of the false orbits. 
Examining the testing sample, we estimate that the contamination of the clean sample is at an averaged level of $\sim10$\%, with a slightly higher level for long-period orbits.

Table~\ref{tab:ROC} lists a few representative threshold values and their corresponding TPR and FPR. These values can be used to select a different SB1 sample with desired sensitivity levels.

\begin{figure*}
	\includegraphics[width=17cm]{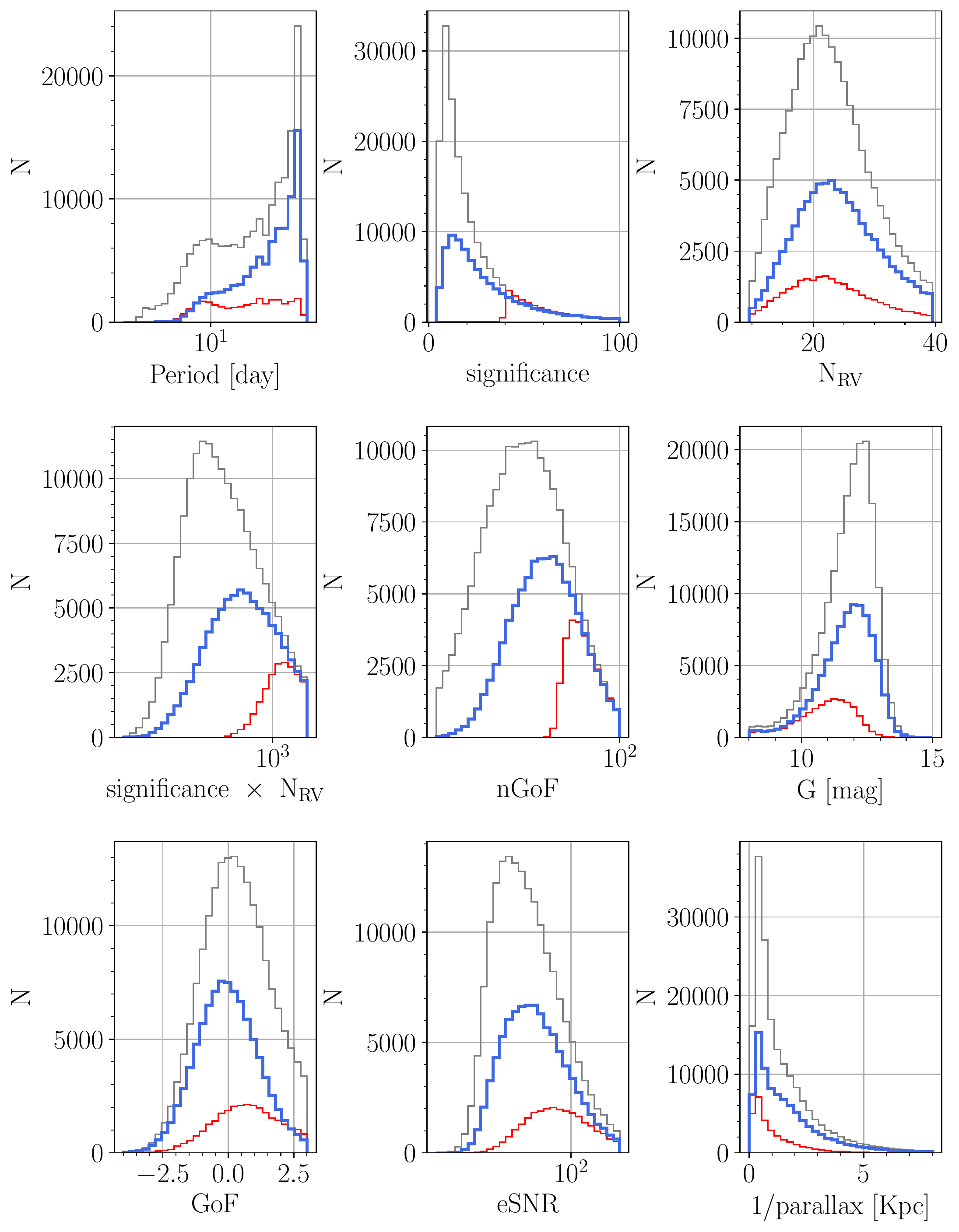}
    \caption{Histogram of the general {\it NSS-SB1} sample ($180\,281$ sources; gray) , the restricted sample of  \texttt{significance > 40} ($28\,774$ sources; red), and our clean \Nclean sources (blue) as function of selected \gaia parameters: Period, \texttt{significance}, N$_{\rm RV}$, \texttt{significance} $\times$ N$_{\rm RV}$, nGoF, G magnitude, GoF, eSNR and distance ($1/\mathrm{parallax}$ [Kpc]). }
    \label{fig:ParamsHistComparison}
\end{figure*}

\begin{table}
\caption{TPR and FPR values for different threshold scores of our logistic-regression clustering algorithm.
\label{tab:ROC}}
\begin{center}
\begin{tabular}{r r r }
\hline
TPR & FPR & Threshold \\
\hline
$0.11$ & $0.01$ & $0.146$ \\
$0.16$ & $0.01$ & $0.158$ \\
$0.21$ & $0.02$ & $0.185$ \\
$0.25$ & $0.02$ & $0.201$ \\
$0.31$ & $0.03$ & $0.228$ \\
$0.35$ & $0.03$ & $0.243$ \\
$0.4$ & $0.05$ & $0.267$ \\
$0.45$ & $0.06$ & $0.3$ \\
$0.5$ & $0.09$ & $0.337$ \\
$0.55$ & $0.14$ & $0.386$ \\
$0.6$ & $0.19$ & $0.422$ \\
$0.65$ & $0.24$ & $0.458$ \\
$0.7$ & $0.29$ & $0.493$ \\
$0.75$ & $0.34$ & $0.526$ \\
$0.8$ & $0.45$ & $0.587$ \\
$0.85$ & $0.55$ & $0.643$ \\
$0.9$ & $0.65$ & $0.695$ \\
$0.95$ & $0.76$ & $0.76$ \\

\hline
\end{tabular}
\end{center}
\end{table}

To demonstrate the realization of our method, Fig.~\ref{fig:ParamsHistComparison} presents the distribution of some of the robust parameters listed for the SB1 solutions for 
three samples:
\begin{itemize}
    \item[--] the {\it NSS-SB1} sample with $180\,281$ orbits with all six robust parameters released (gray curve),
    \item[--] the restricted sample of  \texttt{significance > 40} of $28\,774$ orbits (red curve), considered by {\it NSS}, and
    \item[--] our clean sample of \Nclean orbits (blue curve).
\end{itemize}

As can be seen, our clean sample is about half the size of the general {\it NSS-SB1} one, yet larger by a factor of $\sim 3$ than the \texttt{significance > 40} sample. 
Most notably is that the clean sample includes orbits with lower \texttt{significance} and nGoF, as opposed to the  \texttt{significance > 40} sample. In addition, our clean sample seems to be centered around GoF of $0$, as expected by the SB1 solutions \citep{rimoldini22}.

Comparison between the period distribution of our clean sample
and the original {\it NSS-SB1} one
(Fig.~\ref{fig:ParamsHistComparison} upper left panel)
shows that most of the short-period 
orbits were removed, probably because of erroneous identification of short-period modulations (see above).
We also note a partial removal of the long-period orbits. 
As discussed by \cite{rimoldini22}, this could be due to the fact that the long-period binaries have periods similar to the time span of the \gaia data, and therefore the RVs do not cover more than one claimed orbit. 
Note that even our clean sample might include spurious long-term orbits, as the \gaia estimated RV uncertainties, of $\sim 5$ km/s, are comparable to
the RV amplitudes of their orbits, and therefore we could not identify them as erroneous orbits.




\subsection{Eccentricity-Period Diagram}
\label{e-p}

\begin{figure*}
\includegraphics[width=14cm]{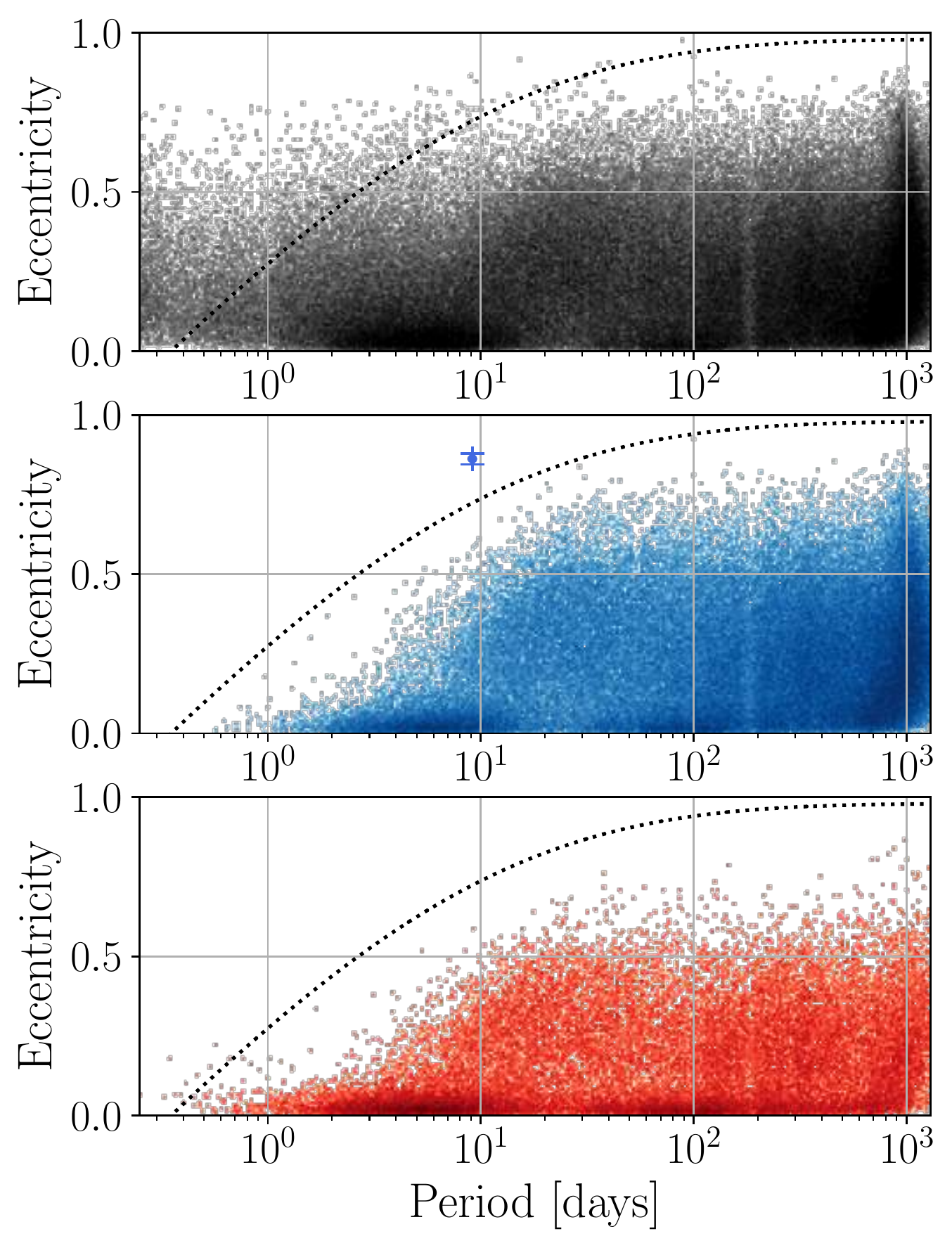}
\caption{Eccentricity vs.~period of the general {\it NSS-SB1} sample ($180\,281$ orbits; upper panel), our clean sample (\Nclean orbits; middle panel) and the restricted sample of  \texttt{significance > 40} ($28\,774$ orbits; bottom panel). A black dotted line marks the observed \citet{Mazeh2008} upper envelope. The blue point in the middle panel marks the only
system in our clean sample found above the expected upper-envelope line.}
    \label{fig:ecc_per}
\end{figure*}

To demonstrate the efficiency of our classification, we consider the eccentricity-period diagram, originally discussed by {\it NSS} and pointed out in the introduction of this paper. As in {\it NSS},  Fig.~\ref{fig:ecc_per} shows that the full sample (upper panel; black) does include many very short-period binaries with high eccentricity (VSPHE), while the restricted sample of \texttt{significance > 40} (lower panel; red) does not show any of those (lower panel). The middle panel (blue) with our clean sample, although much larger than the restricted one, is also clean of these apparently false orbits. 

Out of all binaries of the clean sample, only the eccentricity of one system is significantly higher than the \citet{Mazeh2008} upper-envelope line. The system --- \gaia~DR3~$6727359218822543872$, with reported orbital period $P=9.13712\pm 0.00046$ days and eccentricity $e=0.862\pm0.017$, might be an interesting system to follow up, assuming its orbit is real.

The clean sample displays another interesting feature ---  two dark (dense) concentrations of small-eccentricity orbits around $\sim 5$ and $\sim 90$ days. Those could emanate from circularization processes that were in action for main-sequence (MS) and evolved-stars binaries, respectively. The longer-period concentrations cannot be seen in Fig.~\ref{fig:CornerMS_EV}, probably because that diagram displays only stars with derived masses, and therefore lacks many evolved stars.

\section{Two statistical features of the clean SB1 sample}
\label{sec:statistics}

\begin{figure*}
	\includegraphics[width=12cm]{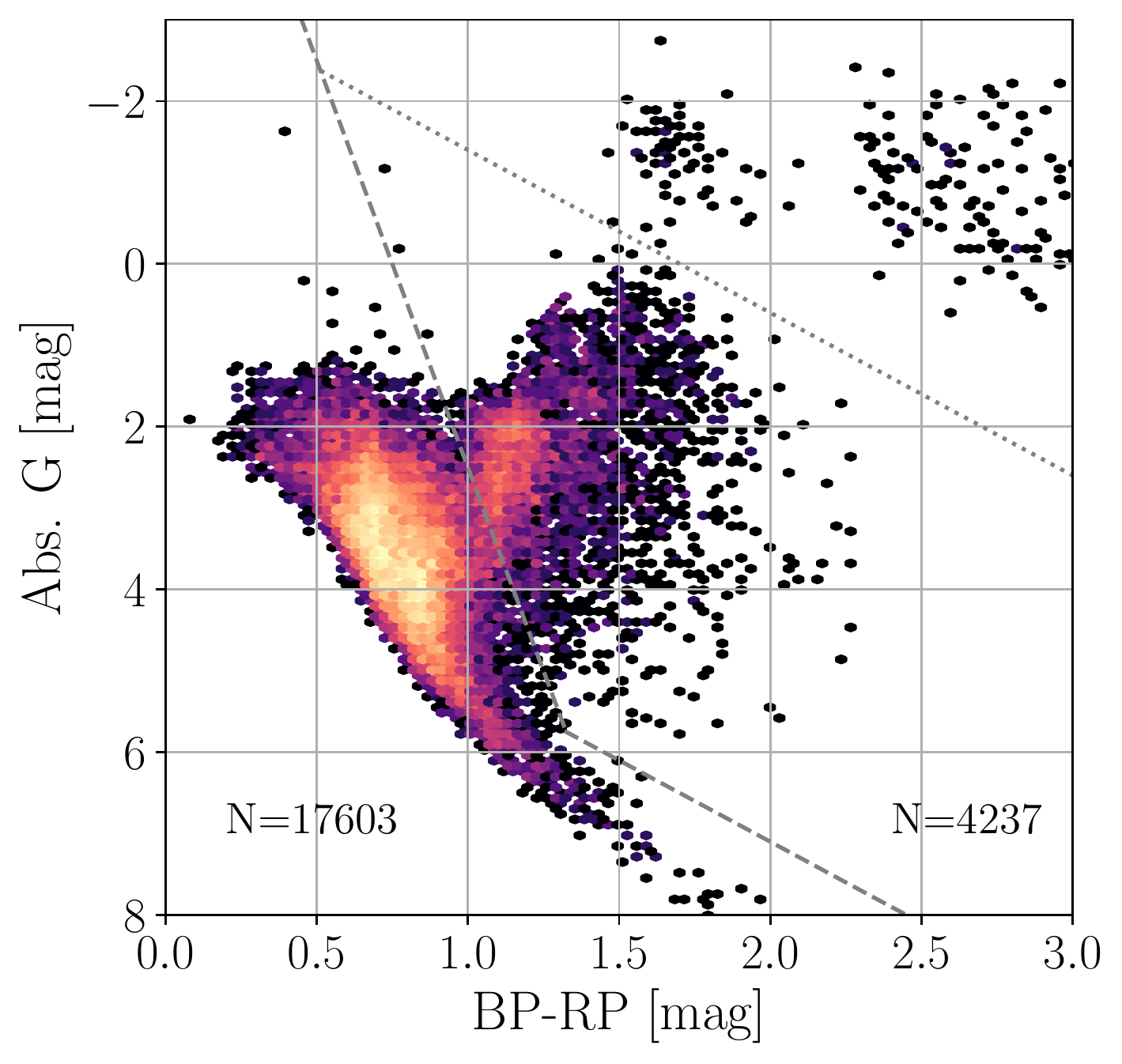}
    \caption{CMD of our clean-sample systems for which a primary-mass estimate is available. Gray dashed line marks the limit that separates our sample of MS and giant stars, while the gray dotted line marks the limit that separates the evolved stars and the long-period variables.}
    \label{fig:CMD}
\end{figure*}

We turn now to the corner plot of the period, eccentricity, and primary mass of the clean sample. 
We wish to derive this plot separately for the MS and the evolved stars, as the correlation between these parameters might be different for the two populations. 
To separate the MS from the giant stars we
plot in Fig.~\ref{fig:CMD} the positions of the binaries of the whole clean sample on the CMD, where the two populations are clearly resolved by:
\begin{equation}
  \mathrm{MS}:\begin{cases}
    -7.5+10(G_{\mathrm{BP}}- G_{\mathrm{RP}}) < G, & \mathrm{if~} G_{\mathrm{BP}}- G_{\mathrm{RP}} \leq 1.32 \ ,\\
    3.1+2(G_{\mathrm{BP}}- G_{\mathrm{RP}}) < G, & \mathrm{if~} G_{\mathrm{BP}}- G_{\mathrm{RP}} > 1.32 \ ,
  \end{cases}
\end{equation}
and the evolved stars from the long-period variables range by
%
\begin{equation}
  \mathrm{LPV}:
    -3.4+2(G_{\mathrm{BP}}- G_{\mathrm{RP}}) > G, \\ .
\end{equation}
%
We get $17\,603$ and $4\,237$ MS and evolved-star orbits, respectively.

\begin{figure*}
	\includegraphics[width=11cm]{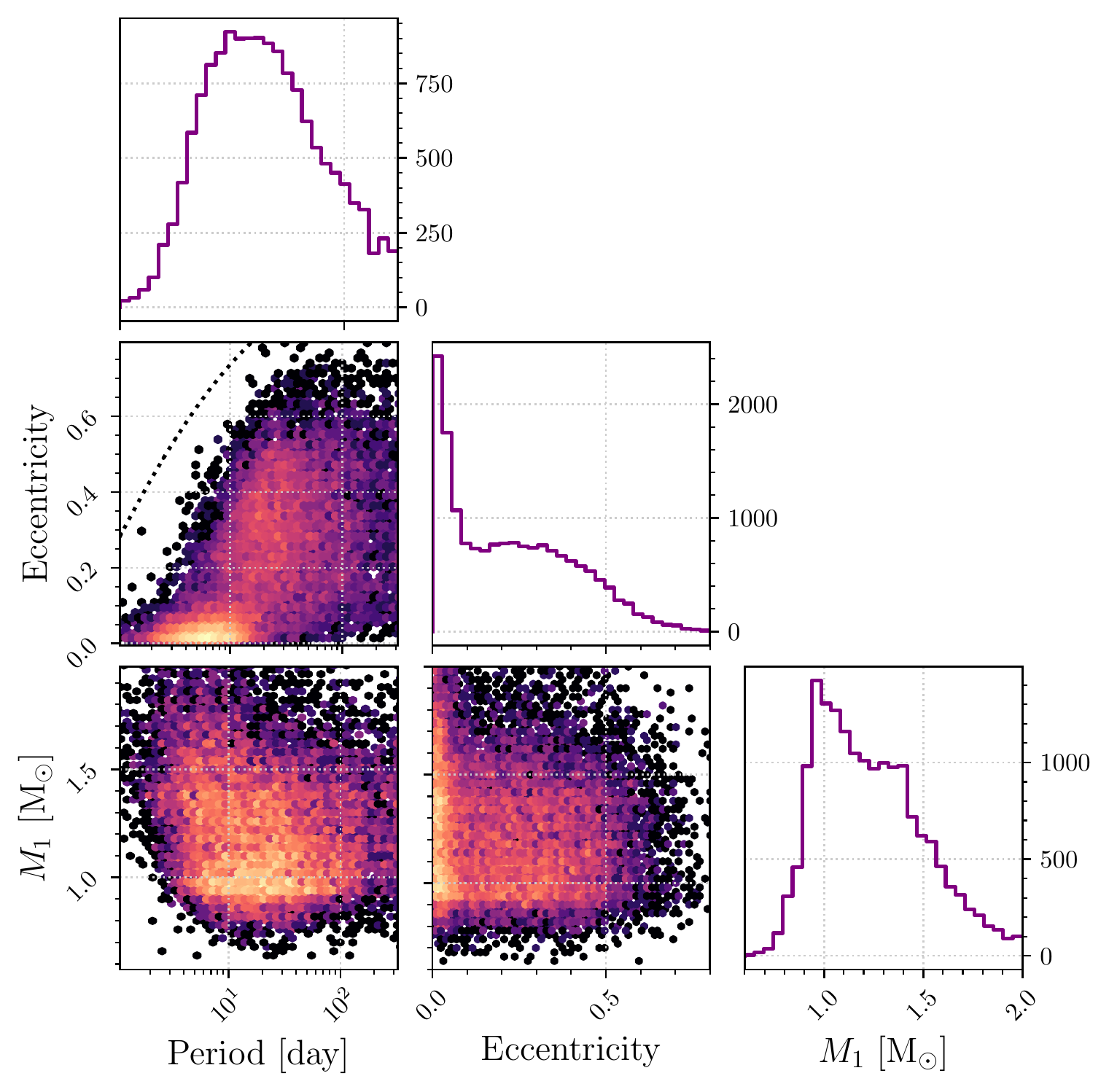}
		\includegraphics[width=11cm]{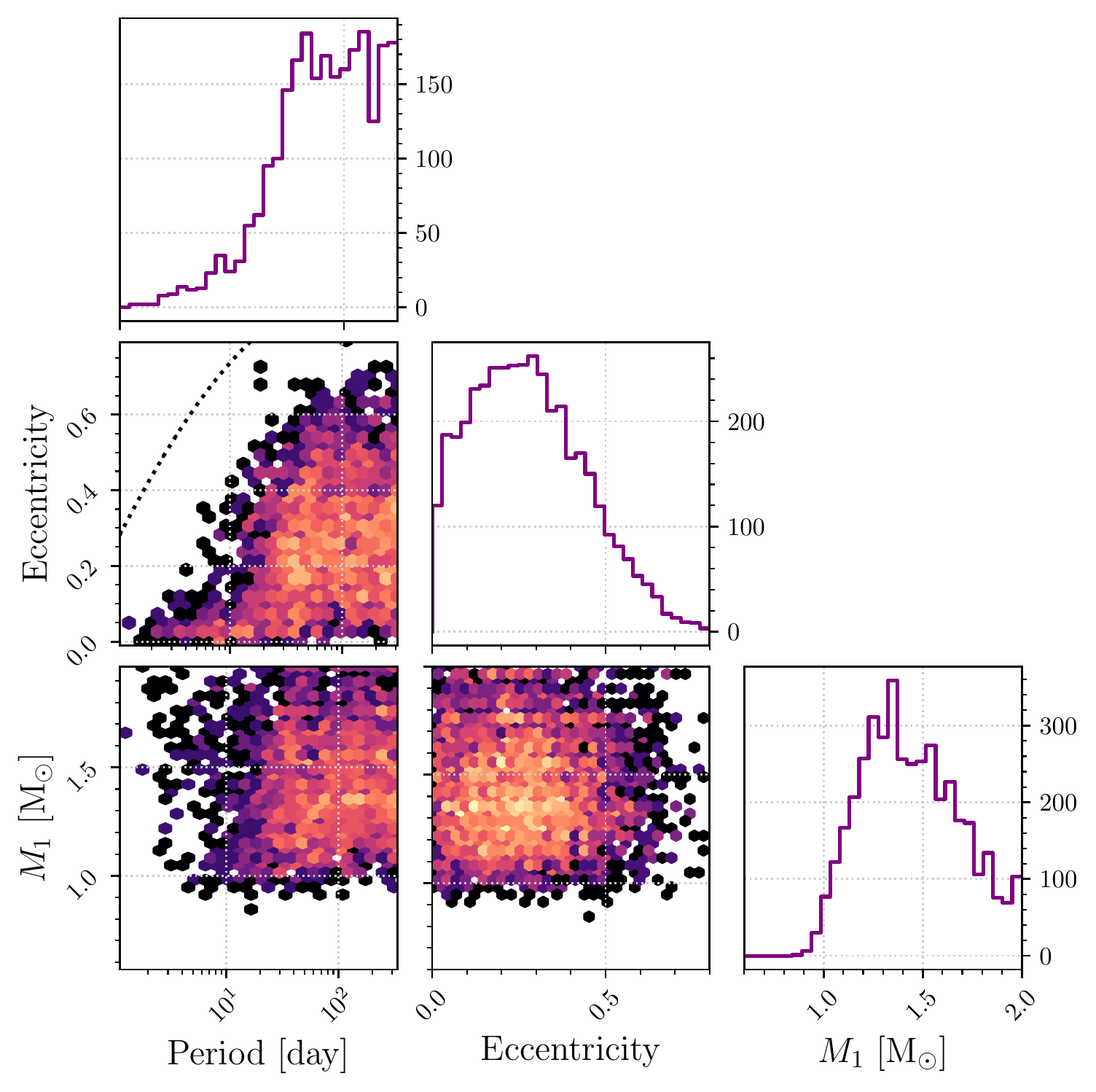}
    \caption{Corner plots of our clean sample of $17\,603$ MS (upper panel) and $4\,237$  evolved (lower panel) stars.}
    \label{fig:CornerMS_EV}
\end{figure*}

%
We then plotted in Fig.~\ref{fig:CornerMS_EV}
the corner plots of the distributions of the period, eccentricity, and primary mass of the MS and evolved stars.  

Two features are emerging:
\begin{itemize}
    \item[--] The MS period-eccentricity diagram suggests two populations: small-eccentricity binaries, in a period range of $2$--$14$ days, and another population with an eccentricity range that fits the \citet{Mazeh2008} curve. The evolved diagram is similar to the second MS population. One can argue that the small eccentricity population is composed of binaries that went through tidal circularization. 
     
    \item[--] A paucity of short-period (up to $5$ day) low-mass (up to $\sim 1.3$ $M_{\odot}$) binaries. 
    
    This feature could come from an observational bias,
    as our classifier is most active at short-period orbits with small RV amplitudes that scale with the 
    primary mass.
\end{itemize}

    In addition, a slight paucity of circular orbits at longer (>$50$ day) periods appears in 
    both the MS and the evolved-star diagrams. 
    The origin of this dearth, discussed in many previous works \citep[][e.g.]{Mathieu94, Jorissen19, PriceWhela20}, may be
    due to an observational bias, as pointed out by \cite{LucySweeney71}.
A spectroscopic binary with a  circular orbit can be found to have small, but nonzero, eccentricity,
    if the SNR of the RV modulation is small.
    In our case, longer-period binaries (>$50$ day) have a lower semi-amplitude modulation and consequently larger RV relative error.

Therefore these features, like many other statistical aspects emerging from the clean catalog, require further study.

\section{Discussion}
\label{sec:summary}

We used the LAMOST DR6 and GALAH DR3 RVs
to validate the orbits of the {\it NSS} \gaia SB1 catalog, finding that $10$--$20$\% of the orbits are probably false.

Our analysis suggests that many of the short-period orbits are false, as was indicated by the eccentricity-period diagram of the original {\it NSS} paper. The reason for this might be the fact that the short-period range of orbits is densely populated by many possible independent false solutions, some of which are consistent with the small number of \gaia RVs. This is especially true for short-period eccentric orbits. Despite cautionary measures by {\it NSS}, some false orbits apparently made their way to the catalog. 
Another cause for false orbits could be the low resolution of \gaia RVS, which did not allow resolving the two components of some double-lined spectroscopic binaries (SB2), introducing wrong solutions. This effect can be stronger in long-period SB2s, for which the RV modulation is smaller, and therefore the two components are not enough separated.

Based on the comparison with LAMOST and GALAH data sets we construct a function that estimates the probability of each of the orbits to be correct, using their published robust parameters.
The function and its values for each of the SB1 are given. Choosing a working point that allows for false-orbit contamination on the order of $10$\%, we were able to put up a selected catalog, consisting of 
\Nclean orbits, for the community to use. 
We have shown that for our particular choice, there are no very-short binaries with high eccentricity in the clean sample.

For three systems with more than $20$ LAMOST RVs measurements, we independently solved for their orbits. 
The three newly derived orbits display different levels of consistency with the \gaia solutions, in agreement with the score derived by our 
classifier for those \gaia orbits.

Obviously, 
the value of the working point, as well as the border lines between false and validated orbits used to construct the probability function, are somewhat arbitrary. Other border lines and working points can be selected, resulting in different clean catalogs, with varying levels of contamination.

The clean sample offers opportunities for studying the short-period binary population. For example, the period-eccentricity diagram of the MS stars of the clean sample shows a sub-sample of small-eccentricity binaries at a period range of $2$--$14$ days, probably composed of binaries that went through tidal circularization. This feature appears in the binaries of evolved stars too, yet in longer periods. Also, the period-primary mass diagram shows 
a paucity of short-period (up to $5$ day) low-mass (up to $\sim 1.3$ $M_{\odot}$) binaries.

These two examples, like many other statistical features hidden in the sample, require further extensive study. In particular, one has to consider observational selection effects that might bias the results, including the specific choices of the clean sample. Furthermore, as shown by {\it NSS}, there is an overlap between the eclipsing and the spectroscopic binaries for the short-period systems, and another overlap between the SB1 systems and the astrometric binaries. A global view of the binary population at hand requires a global analysis of the three \gaia 
{\it NSS} catalogs \citep[see also][]{rimoldini22}, together with other studies, eclipsing binaries found by large photometric surveys \citep[e.g.,][]{Kepler-EB_16,ASAS_EB_06, ZTF_EB_20, ASAS_EB_22} in particular.

\section*{Acknowledgements}
We wish to thank Zephyr Penoyre, the reviewer of this work, for the
thoughtful comments and suggestions that helped us to substantially improve the original manuscript. 
This research was supported by Grant No. 2016069 of the United States-Israel Binational Science Foundation (BSF) and by Grant No. I-1498-303.7/2019 of the German-Israeli Foundation for Scientific Research and Development (GIF) to TM and HWR. The research of SS is supported by a Benoziyo prize postdoctoral fellowship. SD acknowledge support from the National Key R\&D Program of China (No.
2019YFA0405100), the National Natural Science Foundation of China (grant No. 12133005) and the XPLORER PRIZE.

This work has made use of data from the European Space Agency (ESA) mission \gaia (http://www.cosmos.esa.int/gaia), processed
by the Gaia Data Processing and Analysis Consortium (DPAC, http://www.cosmos.esa.int/web/gaia/dpac/consortium). Funding for the DPAC has been provided by national institutions, in particular the institutions participating in the \gaia Multilateral Agreement.

Guoshoujing Telescope (the Large Sky Area Multi-Object Fiber Spectroscopic Telescope LAMOST) is a National Major Scientific Project built by the Chinese Academy of Sciences. Funding for the project has been provided by the National Development and Reform Commission.
LAMOST is operated and managed by the National Astronomical Observatories, Chinese Academy of Sciences. 

This work made use of the Third Data Release of the GALAH Survey \citep{Buder21}. The GALAH Survey is based on data acquired through the Australian Astronomical Observatory, under programs: A/2013B/13 (The GALAH pilot survey); A/2014A/25, A/2015A/19, A2017A/18 (The GALAH survey phase 1); A2018A/18 (Open clusters with HERMES); A2019A/1 (Hierarchical star formation in Ori OB1); A2019A/15 (The GALAH survey phase 2); A/2015B/19, A/2016A/22, A/2016B/10, A/2017B/16, A/2018B/15 (The HERMES-TESS program); and A/2015A/3, A/2015B/1, A/2015B/19, A/2016A/22, A/2016B/12, A/2017A/14 (The HERMES K2-follow-up program). We acknowledge the traditional owners of the land on which the AAT stands, the Gamilaraay people, and pay our respects to elders past and present. This paper includes data that has been provided by AAO Data Central (datacentral.org.au).

This research made use of ASTROPY,\footnote{http://www.astropy.org} a community-developed core Python package for Astronomy \citep{astropy13, astropy18}, 
TOPCAT tool, described in \citep{Taylor05},
and the cross-match service provided by CDS, Strasbourg.

\section*{Data Availability}

Data used in this study are available upon request from the corresponding
author.



\bibliographystyle{mnras}
\bibliography{main} 

\begin{thebibliography}{}
\makeatletter
\relax
\def\mn@urlcharsother{\let\do\@makeother \do\$\do\&\do\#\do\^\do\_\do\%\do\~}
\def\mn@doi{\begingroup\mn@urlcharsother \@ifnextchar [ {\mn@doi@}
  {\mn@doi@[]}}
\def\mn@doi@[#1]#2{\def\@tempa{#1}\ifx\@tempa\@empty \href
  {http://dx.doi.org/#2} {doi:#2}\else \href {http://dx.doi.org/#2} {#1}\fi
  \endgroup}
\def\mn@eprint#1#2{\mn@eprint@#1:#2::\@nil}
\def\mn@eprint@arXiv#1{\href {http://arxiv.org/abs/#1} {{\tt arXiv:#1}}}
\def\mn@eprint@dblp#1{\href {http://dblp.uni-trier.de/rec/bibtex/#1.xml}
  {dblp:#1}}
\def\mn@eprint@#1:#2:#3:#4\@nil{\def\@tempa {#1}\def\@tempb {#2}\def\@tempc
  {#3}\ifx \@tempc \@empty \let \@tempc \@tempb \let \@tempb \@tempa \fi \ifx
  \@tempb \@empty \def\@tempb {arXiv}\fi \@ifundefined
  {mn@eprint@\@tempb}{\@tempb:\@tempc}{\expandafter \expandafter \csname
  mn@eprint@\@tempb\endcsname \expandafter{\@tempc}}}

\bibitem[\protect\citeauthoryear{{Anguiano} et~al.,}{{Anguiano}
  et~al.}{2018}]{Anguiano18}
{Anguiano} B.,  et~al., 2018, \mn@doi [\aap] {10.1051/0004-6361/201833387},
  \href {https://ui.adsabs.harvard.edu/abs/2018A&A...620A..76A} {620, A76}

\bibitem[\protect\citeauthoryear{{Astropy Collaboration} et~al.,}{{Astropy
  Collaboration} et~al.}{2013}]{astropy13}
{Astropy Collaboration} et~al., 2013, \mn@doi [\aap]
  {10.1051/0004-6361/201322068}, \href
  {http://adsabs.harvard.edu/abs/2013A%26A...558A..33A} {558, A33}

\bibitem[\protect\citeauthoryear{{Astropy Collaboration} et~al.,}{{Astropy
  Collaboration} et~al.}{2018}]{astropy18}
{Astropy Collaboration} et~al., 2018, \mn@doi [\aj] {10.3847/1538-3881/aabc4f},
  \href {https://ui.adsabs.harvard.edu/abs/2018AJ....156..123A} {156, 123}

\bibitem[\protect\citeauthoryear{{Barker}}{{Barker}}{2022}]{barker22}
{Barker} A.~J.,  2022, \mn@doi [\apjl] {10.3847/2041-8213/ac5b63}, \href
  {https://ui.adsabs.harvard.edu/abs/2022ApJ...927L..36B} {927, L36}

\bibitem[\protect\citeauthoryear{{Bate} \& {Bonnell}}{{Bate} \&
  {Bonnell}}{1997}]{bate97}
{Bate} M.~R.,  {Bonnell} I.~A.,  1997, \mn@doi [\mnras]
  {10.1093/mnras/285.1.33}, \href
  {http://adsabs.harvard.edu/abs/1997MNRAS.285...33B} {285, 33}

\bibitem[\protect\citeauthoryear{{Bate}, {Bonnell}  \& {Bromm}}{{Bate}
  et~al.}{2002}]{bate02}
{Bate} M.~R.,  {Bonnell} I.~A.,   {Bromm} V.,  2002, \mn@doi [\mnras]
  {10.1046/j.1365-8711.2002.05775.x}, \href
  {https://ui.adsabs.harvard.edu/abs/2002MNRAS.336..705B} {336, 705}

\bibitem[\protect\citeauthoryear{Bishop}{Bishop}{2006}]{Bishop2006}
Bishop C.~M.,  2006, Pattern Recognition and Machine Learning.
Springer, \url {http://research.microsoft.com/en-us/um/people/cmbishop/prml/}

\bibitem[\protect\citeauthoryear{{Blomme} et~al.,}{{Blomme}
  et~al.}{2022}]{RVS_II_22}
{Blomme} R.,  et~al., 2022, arXiv e-prints, \href
  {https://ui.adsabs.harvard.edu/abs/2022arXiv220605486B} {p. arXiv:2206.05486}

\bibitem[\protect\citeauthoryear{{Boffin}}{{Boffin}}{2012}]{boffin12}
{Boffin} H.~M.~J.,  2012, in {Arenou} F.,  {Hestroffer} D.,  eds, Orbital
  Couples: Pas de Deux in the Solar System and the Milky Way. pp 41--44

\bibitem[\protect\citeauthoryear{{Boffin}}{{Boffin}}{2015}]{boffin15}
{Boffin} H.~M.~J.,  2015, \mn@doi [\aap] {10.1051/0004-6361/201525762}, \href
  {http://adsabs.harvard.edu/abs/2015A%26A...575L..13B} {575, L13}

\bibitem[\protect\citeauthoryear{{Buder} et~al.,}{{Buder}
  et~al.}{2021}]{Buder21}
{Buder} S.,  et~al., 2021, \mn@doi [\mnras] {10.1093/mnras/stab1242}, \href
  {https://ui.adsabs.harvard.edu/abs/2021MNRAS.506..150B} {506, 150}

\bibitem[\protect\citeauthoryear{{Chen}, {Wang}, {Deng}, {de Grijs}, {Yang}  \&
  {Tian}}{{Chen} et~al.}{2020}]{ZTF_EB_20}
{Chen} X.,  {Wang} S.,  {Deng} L.,  {de Grijs} R.,  {Yang} M.,   {Tian} H.,
  2020, \mn@doi [\apjs] {10.3847/1538-4365/ab9cae}, \href
  {https://ui.adsabs.harvard.edu/abs/2020ApJS..249...18C} {249, 18}

\bibitem[\protect\citeauthoryear{Cui et~al.,}{Cui et~al.}{2012}]{cui12}
Cui X.-Q.,  et~al., 2012, RAA, 12, 1197

\bibitem[\protect\citeauthoryear{{Duch{\^e}ne} \& {Kraus}}{{Duch{\^e}ne} \&
  {Kraus}}{2013}]{duchene13}
{Duch{\^e}ne} G.,  {Kraus} A.,  2013, \mn@doi [\araa]
  {10.1146/annurev-astro-081710-102602}, \href
  {https://ui.adsabs.harvard.edu/abs/2013ARA&A..51..269D} {51, 269}

\bibitem[\protect\citeauthoryear{Fawcett}{Fawcett}{2006}]{fawcett06}
Fawcett T.,  2006, \mn@doi [Pattern Recognition Letters]
  {https://doi.org/10.1016/j.patrec.2005.10.010}, 27, 861

\bibitem[\protect\citeauthoryear{{Gaia Collaboration} et~al.,}{{Gaia
  Collaboration} et~al.}{2022}]{NSS}
{Gaia Collaboration} et~al., 2022, arXiv e-prints, \href
  {https://ui.adsabs.harvard.edu/abs/2022arXiv220605595G} {p. arXiv:2206.05595}

\bibitem[\protect\citeauthoryear{{Harada}, {Hirano}, {Machida}  \&
  {Hosokawa}}{{Harada} et~al.}{2021}]{harada21}
{Harada} N.,  {Hirano} S.,  {Machida} M.~N.,   {Hosokawa} T.,  2021, \mn@doi
  [\mnras] {10.1093/mnras/stab2780}, \href
  {https://ui.adsabs.harvard.edu/abs/2021MNRAS.508.3730H} {508, 3730}

\bibitem[\protect\citeauthoryear{{Jorissen}, {Frankowski}, {Famaey}  \& {van
  Eck}}{{Jorissen} et~al.}{2009}]{jorissen09}
{Jorissen} A.,  {Frankowski} A.,  {Famaey} B.,   {van Eck} S.,  2009, \mn@doi
  [\aap] {10.1051/0004-6361/200810703}, \href
  {https://ui.adsabs.harvard.edu/abs/2009A&A...498..489J} {498, 489}

\bibitem[\protect\citeauthoryear{{Jorissen}, {Boffin}, {Karinkuzhi}, {Van Eck},
  {Escorza}, {Shetye}  \& {Van Winckel}}{{Jorissen} et~al.}{2019}]{Jorissen19}
{Jorissen} A.,  {Boffin} H.~M.~J.,  {Karinkuzhi} D.,  {Van Eck} S.,  {Escorza}
  A.,  {Shetye} S.,   {Van Winckel} H.,  2019, \mn@doi [\aap]
  {10.1051/0004-6361/201834630}, \href
  {https://ui.adsabs.harvard.edu/abs/2019A&A...626A.127J} {626, A127}

\bibitem[\protect\citeauthoryear{{Katz} et~al.,}{{Katz}
  et~al.}{2022}]{DR3_Katz}
{Katz} D.,  et~al., 2022, arXiv e-prints, \href
  {https://ui.adsabs.harvard.edu/abs/2022arXiv220605902K} {p. arXiv:2206.05902}

\bibitem[\protect\citeauthoryear{{Kirk} et~al.,}{{Kirk}
  et~al.}{2016}]{Kepler-EB_16}
{Kirk} B.,  et~al., 2016, \mn@doi [\aj] {10.3847/0004-6256/151/3/68}, \href
  {https://ui.adsabs.harvard.edu/abs/2016AJ....151...68K} {151, 68}

\bibitem[\protect\citeauthoryear{{Lucy} \& {Sweeney}}{{Lucy} \&
  {Sweeney}}{1971}]{LucySweeney71}
{Lucy} L.~B.,  {Sweeney} M.~A.,  1971, \mn@doi [\aj] {10.1086/111159}, \href
  {https://ui.adsabs.harvard.edu/abs/1971AJ.....76..544L} {76, 544}

\bibitem[\protect\citeauthoryear{{Mathieu}}{{Mathieu}}{1994}]{Mathieu94}
{Mathieu} R.~D.,  1994, \mn@doi [\araa] {10.1146/annurev.aa.32.090194.002341},
  \href {https://ui.adsabs.harvard.edu/abs/1994ARA&A..32..465M} {32, 465}

\bibitem[\protect\citeauthoryear{{Mazeh}}{{Mazeh}}{2008}]{Mazeh2008}
{Mazeh} T.,  2008, in {Goupil} M.~J.,  {Zahn} J.~P.,  eds,  EAS Publications
  Series Vol. 29, EAS Publications Series. pp 1--65 (\mn@eprint {arXiv}
  {0801.0134}), \mn@doi{10.1051/eas:0829001}

\bibitem[\protect\citeauthoryear{{Mazeh} \& {Goldberg}}{{Mazeh} \&
  {Goldberg}}{1992}]{mg92}
{Mazeh} T.,  {Goldberg} D.,  1992, \mn@doi [\apj] {10.1086/171611}, \href
  {http://adsabs.harvard.edu/abs/1992ApJ...394..592M} {394, 592}

\bibitem[\protect\citeauthoryear{{Meibom} \& {Mathieu}}{{Meibom} \&
  {Mathieu}}{2005}]{meibom05}
{Meibom} S.,  {Mathieu} R.~D.,  2005, \mn@doi [\apj] {10.1086/427082}, \href
  {https://ui.adsabs.harvard.edu/abs/2005ApJ...620..970M} {620, 970}

\bibitem[\protect\citeauthoryear{{Moe} \& {Di Stefano}}{{Moe} \& {Di
  Stefano}}{2017}]{moe17}
{Moe} M.,  {Di Stefano} R.,  2017, \mn@doi [\apjs] {10.3847/1538-4365/aa6fb6},
  \href {http://adsabs.harvard.edu/abs/2017ApJS..230...15M} {230, 15}

\bibitem[\protect\citeauthoryear{{Paczy{\'n}ski}, {Szczygie{\l}}, {Pilecki}  \&
  {Pojma{\'n}ski}}{{Paczy{\'n}ski} et~al.}{2006}]{ASAS_EB_06}
{Paczy{\'n}ski} B.,  {Szczygie{\l}} D.~M.,  {Pilecki} B.,   {Pojma{\'n}ski} G.,
   2006, \mn@doi [\mnras] {10.1111/j.1365-2966.2006.10223.x}, \href
  {https://ui.adsabs.harvard.edu/abs/2006MNRAS.368.1311P} {368, 1311}

\bibitem[\protect\citeauthoryear{{Petrosky}, {Hwang}, {Zakamska}, {Chandra}  \&
  {Hill}}{{Petrosky} et~al.}{2021}]{Petrosky21}
{Petrosky} E.,  {Hwang} H.-C.,  {Zakamska} N.~L.,  {Chandra} V.,   {Hill}
  M.~J.,  2021, \mn@doi [\mnras] {10.1093/mnras/stab592}, \href
  {https://ui.adsabs.harvard.edu/abs/2021MNRAS.503.3975P} {503, 3975}

\bibitem[\protect\citeauthoryear{{Pourbaix} et~al.,}{{Pourbaix}
  et~al.}{2007}]{SB9}
{Pourbaix} D.,  et~al., 2007, VizieR Online Data Catalog, \href
  {https://ui.adsabs.harvard.edu/abs/2007yCat....102020P} {p. B/sb9}

\bibitem[\protect\citeauthoryear{{Price-Whelan} \& {Goodman}}{{Price-Whelan} \&
  {Goodman}}{2018}]{price-whelan18}
{Price-Whelan} A.~M.,  {Goodman} J.,  2018, \mn@doi [\apj]
  {10.3847/1538-4357/aae264}, \href
  {https://ui.adsabs.harvard.edu/abs/2018ApJ...867....5P} {867, 5}

\bibitem[\protect\citeauthoryear{{Price-Whelan} et~al.,}{{Price-Whelan}
  et~al.}{2020a}]{Price-Whelan20}
{Price-Whelan} A.~M.,  et~al., 2020a, \mn@doi [\apj]
  {10.3847/1538-4357/ab8acc}, \href
  {https://ui.adsabs.harvard.edu/abs/2020ApJ...895....2P} {895, 2}

\bibitem[\protect\citeauthoryear{{Price-Whelan} et~al.,}{{Price-Whelan}
  et~al.}{2020b}]{PriceWhela20}
{Price-Whelan} A.~M.,  et~al., 2020b, \mn@doi [\apj]
  {10.3847/1538-4357/ab8acc}, \href
  {https://ui.adsabs.harvard.edu/abs/2020ApJ...895....2P} {895, 2}

\bibitem[\protect\citeauthoryear{{Raghavan} et~al.,}{{Raghavan}
  et~al.}{2010}]{raghavan10}
{Raghavan} D.,  et~al., 2010, \mn@doi [\apjs] {10.1088/0067-0049/190/1/1},
  \href {http://adsabs.harvard.edu/abs/2010ApJS..190....1R} {190, 1}

\bibitem[\protect\citeauthoryear{{Recio-Blanco} et~al.,}{{Recio-Blanco}
  et~al.}{2022}]{RVS_I_22}
{Recio-Blanco} A.,  et~al., 2022, arXiv e-prints, \href
  {https://ui.adsabs.harvard.edu/abs/2022arXiv220605541R} {p. arXiv:2206.05541}

\bibitem[\protect\citeauthoryear{{Rimoldini} et~al.,}{{Rimoldini}
  et~al.}{2022}]{rimoldini22}
{Rimoldini} L.,  et~al., 2022, {Gaia DR3 documentation Chapter 10:
  Variability}, Gaia DR3 documentation.

\bibitem[\protect\citeauthoryear{{Rowan} et~al.,}{{Rowan}
  et~al.}{2022}]{ASAS_EB_22}
{Rowan} D.~M.,  et~al., 2022, arXiv e-prints, \href
  {https://ui.adsabs.harvard.edu/abs/2022arXiv220505687R} {p. arXiv:2205.05687}

\bibitem[\protect\citeauthoryear{{Shahaf} \& {Mazeh}}{{Shahaf} \&
  {Mazeh}}{2019}]{shahaf19}
{Shahaf} S.,  {Mazeh} T.,  2019, \mn@doi [\mnras] {10.1093/mnras/stz1517},
  \href {https://ui.adsabs.harvard.edu/abs/2019MNRAS.487.3356S} {487, 3356}

\bibitem[\protect\citeauthoryear{{Shahaf}, {Mazeh}  \& {Faigler}}{{Shahaf}
  et~al.}{2017}]{shahaf17}
{Shahaf} S.,  {Mazeh} T.,   {Faigler} S.,  2017, \mn@doi [\mnras]
  {10.1093/mnras/stx2257}, \href
  {https://ui.adsabs.harvard.edu/abs/2017MNRAS.472.4497S} {472, 4497}

\bibitem[\protect\citeauthoryear{{Taylor}}{{Taylor}}{2005}]{Taylor05}
{Taylor} M.~B.,  2005, in {Shopbell} P.,  {Britton} M.,   {Ebert} R.,  eds,
  Astronomical Society of the Pacific Conference Series Vol. 347, Astronomical
  Data Analysis Software and Systems XIV. p.~29

\bibitem[\protect\citeauthoryear{{Terquem} \& {Martin}}{{Terquem} \&
  {Martin}}{2021}]{terquem21}
{Terquem} C.,  {Martin} S.,  2021, \mn@doi [\mnras] {10.1093/mnras/stab2322},
  \href {https://ui.adsabs.harvard.edu/abs/2021MNRAS.507.4165T} {507, 4165}

\bibitem[\protect\citeauthoryear{{Troup} et~al.,}{{Troup}
  et~al.}{2016}]{troup16}
{Troup} N.~W.,  et~al., 2016, \mn@doi [\aj] {10.3847/0004-6256/151/3/85}, \href
  {https://ui.adsabs.harvard.edu/abs/2016AJ....151...85T} {151, 85}

\bibitem[\protect\citeauthoryear{{Van Eylen}, {Winn}  \& {Albrecht}}{{Van
  Eylen} et~al.}{2016}]{winn16}
{Van Eylen} V.,  {Winn} J.~N.,   {Albrecht} S.,  2016, \mn@doi [\apj]
  {10.3847/0004-637X/824/1/15}, \href
  {https://ui.adsabs.harvard.edu/abs/2016ApJ...824...15V} {824, 15}

\bibitem[\protect\citeauthoryear{{Verbunt} \& {Phinney}}{{Verbunt} \&
  {Phinney}}{1995}]{verbunt95}
{Verbunt} F.,  {Phinney} E.~S.,  1995, \aap, \href
  {https://ui.adsabs.harvard.edu/abs/1995A&A...296..709V} {296, 709}

\makeatother
\end{thebibliography}



\appendix

\section{Three LAMOST orbits --- comparison with the {\it NSS-SB1} solutions}
\label{sec:three-solutions}

For three systems with more than 20 LAMOST RVs measurements, we independently solved for the orbits. 
We list in Table \ref{tab:LASMOT+20sources} our LAMOST elements of these three systems together with those of \gaia.

\begin{figure}
	\includegraphics[width=\columnwidth]{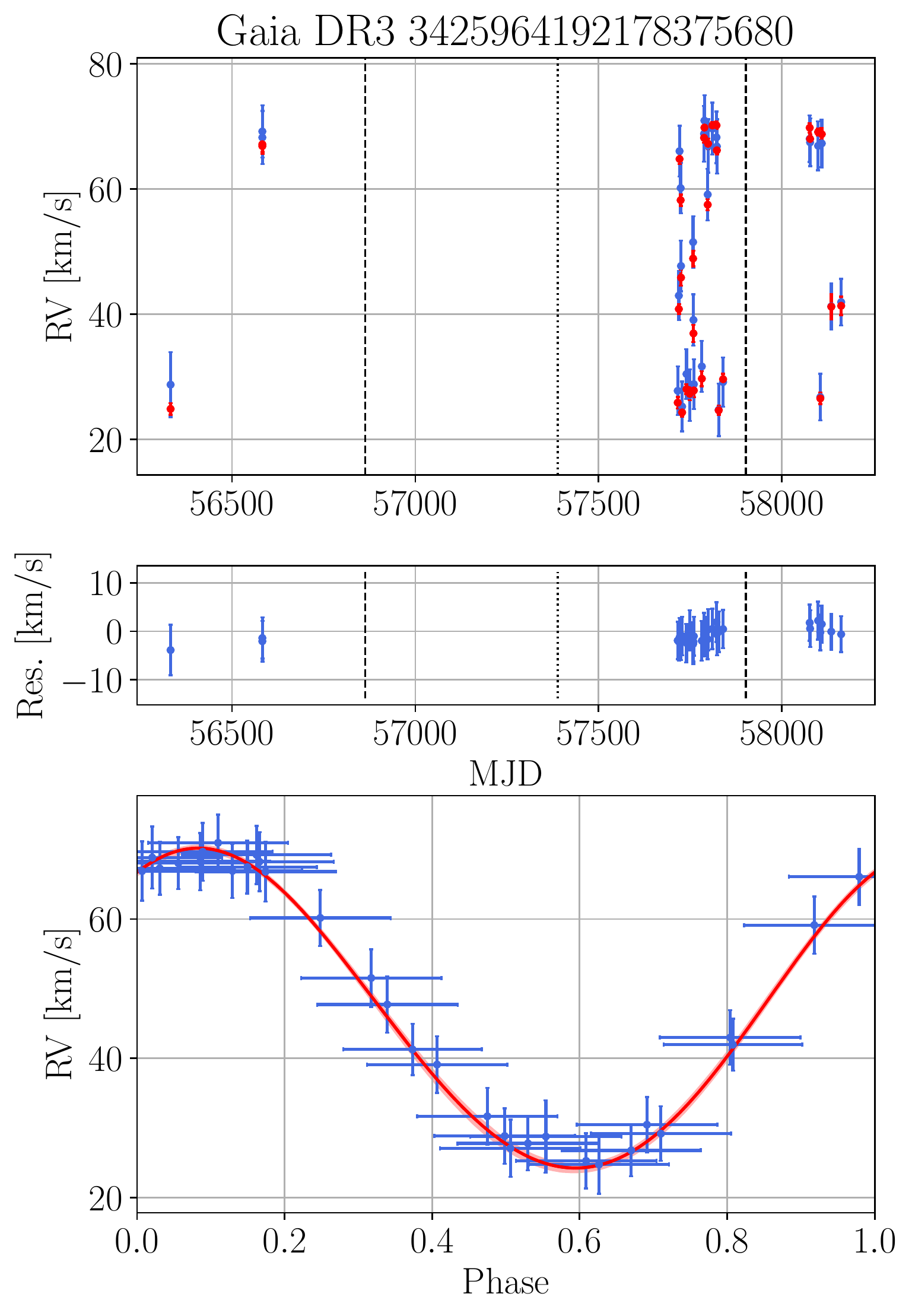}
    \caption{ \gaia model of 3425964192178375680 with the LAMOST RVs at \textit{Gaia} phases. Upper panel displays the reported RVs of LAMOST (blue points) and expected \gaia (red points) as a function of time. Dashed vertical lines mark the time span of \gaia DR3 observation while dotted vertical line marks the system's periastron time reported by \gaia. Middle panel displays the residuals on the same time scale. Bottom panel presents the \gaia phase folded model (red curve) and the LAMOST RVs (blue points) with their uncertainties.}
    \label{fig:GaiaModel_3425964192178375680}
\end{figure}

\begin{figure}
	\includegraphics[width=\columnwidth]{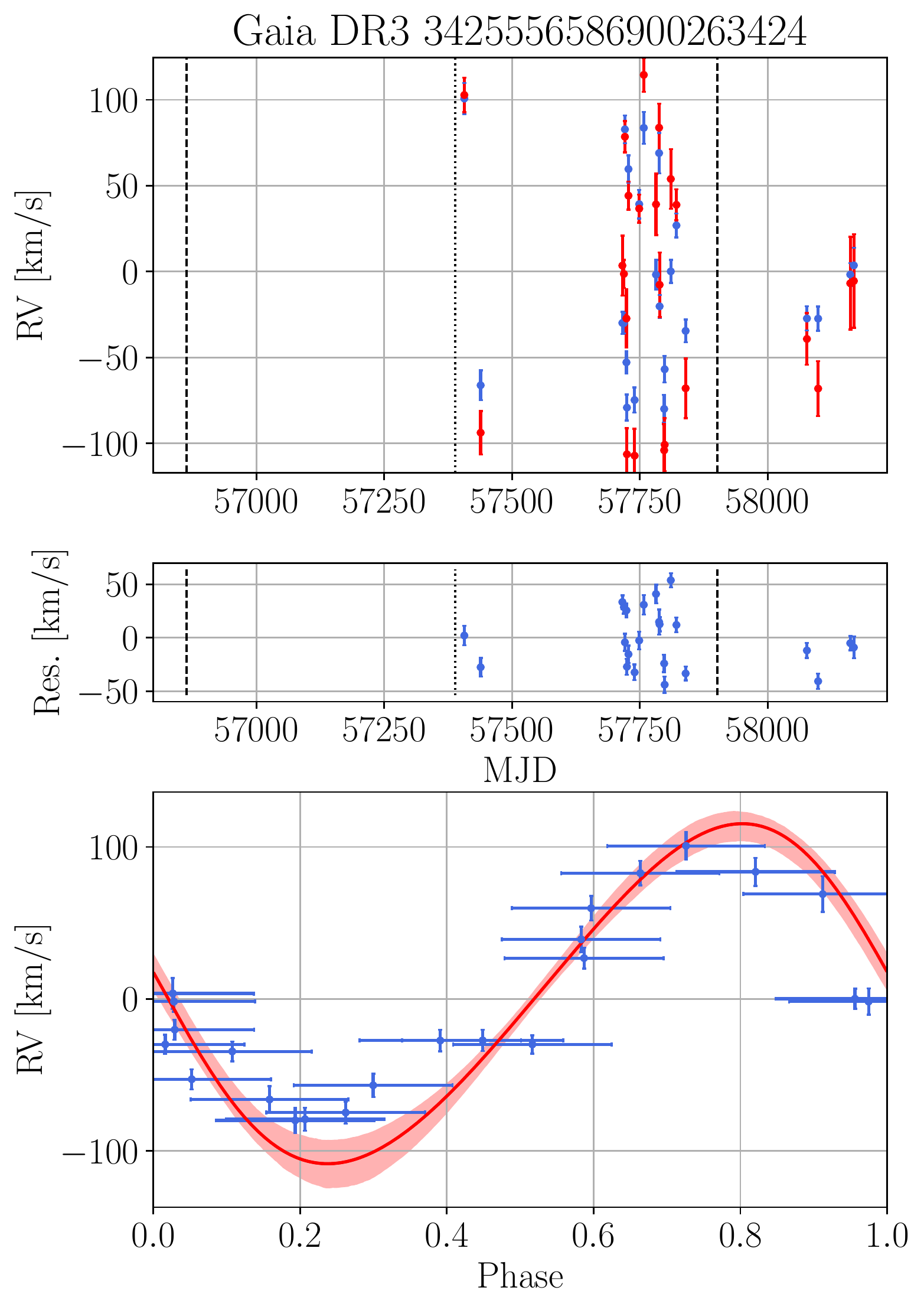}
    \caption{ 
    \gaia model of 3425556586900263424 with the LAMOST RVs at \textit{Gaia} phases (see Fig.~\ref{fig:GaiaModel_3425964192178375680}).}
    \label{fig:GaiaModel_3425556586900263424}
\end{figure}

\begin{figure}
	\includegraphics[width=\columnwidth]{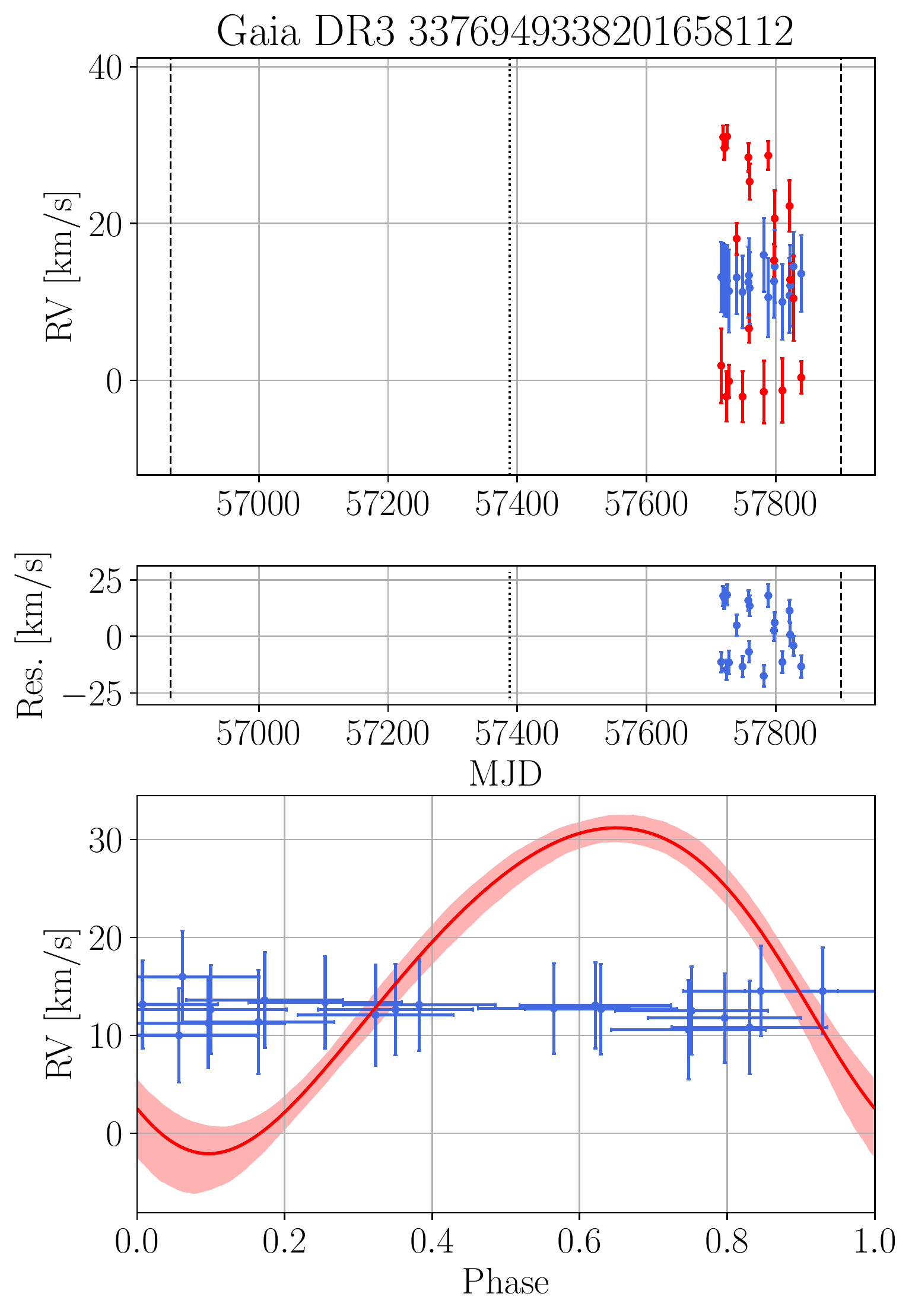}
    \caption{  
    \gaia model of 3376949338201658112 with the LAMOST RVs at \textit{Gaia} phases (see 
    Fig.~\ref{fig:GaiaModel_3425964192178375680}).}
    \label{fig:GaiaModel_3376949338201658112}
\end{figure}

In Fig.~\ref{fig:GaiaModel_3425964192178375680},  \ref{fig:GaiaModel_3425556586900263424} and \ref{fig:GaiaModel_3376949338201658112} we plotted the \gaia folded model, together with the LAMOST RVs, at phases derived with the LAMOST epochs and the period and phase of \gaia.  As can be seen, each one of the three systems demonstrates a different level of consistency with the \gaia solutions, as well as with the score found by our 
classifier. 


\begin{table*}
\caption{The adopted Keplerian-orbit parameters of the three LAMOST sources with more than $20$ RV measurements, compared with the Gaia elements.
\label{tab:LASMOT+20sources}}
\begin{center}
\begin{tabular}{rr rrrrrr}
\hline
{} & & \multicolumn{2}{r}{\gaia DR3 $3425964192178375680~$}& \multicolumn{2}{r}{\textit{Gaia} DR3 $3425556586900263424~$} & \multicolumn{2}{r}{\gaia DR3 $3376949338201658112~$} \\
Parameter & Units  & \gaia & LAMOST  & \textit{Gaia} & LAMOST  & \gaia & LAMOST  \\
\hline
N$_{\rm RV}$ &    &  $16$   & $31$ & $15$ & $23$ & $13$& $20$ \\

Time span &  day  &  $953.55$   & $1828.89$  &  $966.62$   & $761.89$  &  $965.87$   & $123.73$\\

nGoF &    &  $42.41$   & ---  &  $16.14$   & ---  &  $15.16$   & ---
\\
\hline
Period & day   &  $11.0672\pm 0.0020$   & $11.0679\pm 0.0064$  &  $0.4667616 \pm 7.2\times 10^{-6}$ & $0.877526\pm0.000021$  & $0.657582 \pm 2.2\times 10^{-5}$ & ---  \\

K & km/s   &  $22.9\pm 0.6$   &  $22.2 \pm 1.0$ &  $111.9 \pm 8.2$ & $79.9\pm2.8$ &$15.9\pm 1.1$ & ---  \\

e &    &  $0.033\pm 0.023$   & $0.043\pm 0.033$  & $0.103 \pm 0.062$  & $0.053 \pm 0.031$ & $0.118\pm 0.111$ &  --- \\

$T_p$ & day   &  $57388.93\pm 1.04$   & $	57392.69\pm 0.30$  &  $57388.539 \pm 0.049$   & $57387.519\pm0.014$ & $57388.266 \pm 0.061$ &  --- \\

$\omega$ &  deg   &  $328 \pm 34$   & $ 338.26 \pm109 $ & $82.7 \pm 37.4$& $41.5 \pm 51.8$&   $136.3\pm34.6$ & ---  \\

$\gamma$ & km/s  &  $46.57\pm 0.34$   & $47.66\pm 0.73$  & $1.9 \pm 5.1$   &  $-7.7 \pm 1.3$ &$15.9 \pm 1.1$ & $12.7 \pm 4.7$  \\

\hline
Score &    &  $0.63$   & ---  &  $0.28$   & ---  &  $0.16$   & ---
\\
\hline
\end{tabular}
\end{center}
\end{table*}

\bsp	
\label{lastpage}
\end{document}